\documentclass[12pt,a4paper]{article}
\usepackage{graphics}
\usepackage{graphicx}
\begin{document}
\newcommand{\fig}{} 
\title{Fractal and Multifractal Time Series}
\author{Jan W. Kantelhardt\\
Institute of Physics, Martin-Luther-University\\
Halle-Wittenberg, 06099 Halle, Germany}
\maketitle
\tableofcontents

\section*{Glossary}
\begin{itemize}
\item{\bf Time series}:  One dimensional array of numbers $(x_i)$, $i=1,\ldots,N$, 
 representing values of an observable $x$ usually measured equidistant (or nearly 
 equidistant) in time.  
\item{\bf Complex system}: A system consisting of many non-linearly interacting 
 components. It cannot be split into simpler sub-systems without tampering the 
 dynamical properties.
\item{\bf Scaling law}: A power law with a scaling exponent (e.~g. $\alpha$) 
 describing the behaviour of a quantity $F$ (e.~g., fluctuation, spectral power) 
 as function of a scale parameter $s$ (e.~g., time scale, frequency) at least 
 asymptotically: $F(s) \sim s^\alpha$.  The power law should be valid for
 a large range of $s$ values, e.~g., at least for one order of magnitude.
\item{\bf Fractal system}:  A system characterised by a scaling law with a fractal,
 i.~e., non-integer exponent.  Fractal systems are characterised by self-similarity,
 i.~e., a magnification of a small part is statstically equivalent to the whole.
\item{\bf Self-affine system}:  Generalization of a fractal system, where different
 magnifications $s$ and $s'=s^H$ have to be used for different directions in 
 order to obtain a statistically equivalent magnification.  The exponent $H$ is 
 called Hurst exponent.  Self-affine time series and time series becoming self-affine
 upon integration are commonly denoted as fractal using a less strict terminology.
\item{\bf Multifractal system}:  A system characterised by scaling laws with an 
 infinite number of different fractal exponents.  The scaling laws must be valid 
 for the same range of the scale parameter.
\item{\bf Crossover}:  Change point in a scaling law, where one scaling exponent 
 applies for small scale parameters and another scaling exponent applies for large
 scale parameters.  The center of the crossover is denoted by its characteristic
 scale parameter $s_\times$ in this article.
\item{\bf Persistence}:  In a persistent time series, a large value is usually 
 (i.~e., with high statistical preference) followed by a large value and a small 
 value is followed by a small value.  A fractal scaling law holds at least for a 
 limited range of scales.
\item{\bf Short-term correlations}:  Correlations that decay sufficiently fast 
 that they can be described by a characteristic correlation time scale; e.~g., 
 exponentially decaying correlations.  A crossover to uncorrelated behaviour is 
 observed on larger scales.
\item{\bf Long-term correlations}:  Correlations that decay sufficiently slow that
 a characteristic correlation time scale cannot be defined; e.~g., power-law 
 correlations with an exponent between 0 and 1.  Power-law scaling is observed on 
 large time scales and asymptotically.  The term long-range correlations should
 be used if the data is not a time series.
\item{\bf Non-stationarities}:  If the mean or the standard deviation of the data 
 values change with time, the weak definition of stationarity is violated. The 
 strong definition of stationarity requires that all moments remain constant, 
 i.~e., the distribution density of the values does not change with time. 
 Non-stationarities like monotonous, periodic, or step-like trends are often 
 caused by external effects.  In a more general sense, changes in the dynamics 
 of the system also represent non-stationarities.
\end{itemize}

\section{Definition of the Subject and Its Importance}

Data series generated by complex systems exhibit fluctuations on a wide range 
of time scales and/or broad distributions of the values.  In both equilibrium 
and non-equilibrium situations, the natural fluctuations are often found to 
follow a scaling relation over several orders of magnitude.  Such scaling laws
allow for a characterisation of the data and the generating complex system by 
fractal (or multifractal) scaling exponents, which can serve as characteristic 
fingerprints of the systems in comparisons with other systems and with models.  
Fractal scaling behaviour has been observed, e.~g., in many data series from 
experimental physics, geophysics, medicine, physiology, and even social sciences. 
Although the underlying causes of the observed fractal scaling are often not 
known in detail, the fractal or multifractal characterisation can be used for 
generating surrogate (test) data, modelling the time series, and deriving 
predictions regarding extreme events or future behaviour.  The main application, 
however, is still the characterisation of different states or phases of the 
complex system based on the observed scaling behaviour.  For example, the health 
status and different physiological states of the human cardiovascular system 
are represented by the fractal scaling behaviour of the time series of intervals 
between successive heartbeats, and the coarsening  dynamics in metal alloys are 
represented by the fractal scaling of the time-dependent speckle intensities 
observed in coherent X-ray spectroscopy.  

In order to observe fractal and multifractal scaling behaviour in time series, 
several tools have been developed.  Besides older techniques assuming stationary
data, there are more recently established methods differentiating truly fractal 
dynamics from fake scaling behaviour caused by non-stationarities in the data.  
In addition, short-term and long-term correlations have to be clearly 
distinguished to show fractal scaling behaviour unambiguously.  This article 
describes several methods originating from Statistical Physics and Applied 
Mathematics, which have been used for fractal and multifractal time series 
analysis in stationary and non-stationary data.

\section{Introduction}

The characterisation and understanding of complex systems is a difficult task,
since they cannot be split into simpler subsystems without tampering the dynamical
properties.  One approach in studying such systems is the recording of long {\it 
time series} of several selected variables (observables), which reflect the state 
of the system in a dimensionally reduced representation.  Some systems are 
characterised by periodic or nearly periodic behaviour, which might be caused by 
oscillatory components or closed-loop regulation chains.  However, in truly 
complex systems such periodic components are usually not limited to one or two 
characteristic frequencies or frequency bands.  They rather extend over a wide 
spectrum, and fluctuations on many time scales as well as broad distributions of 
the values are found.  Often no specific lower frequency limit -- or, equivalently, 
upper characteristic time scale -- can be observed.  In these cases, the dynamics 
can be characterised by {\it scaling laws} which are valid over a wide (possibly 
even unlimited) range of time scales or frequencies; at least over orders of 
magnitude.  Such dynamics are usually denoted as {\it fractal} or {\it multifractal}, 
depending on the question if they are characterised by one scaling exponent or by 
a multitude of scaling exponents.  

The first scientist who applied fractal analysis to natural time series is Benoit 
B. Mandelbrot \cite{mandel68,mandel69,mandel99}, who included early approaches 
by H.E. Hurst regarding hydrological systems \cite{hurst51,hurst65}. For extensive 
introductions describing fractal scaling in complex systems, we refer to 
\cite{feder88,barnsley93,bunde94,jorgensen00,bunde02,kantz03,peitgen04,sornette04}.
In the last decade, fractal and multifractal scaling behaviour has been reported 
in many natural time series generated by complex systems, including
\begin{itemize} 
\item geophysics time series (recordings of temperature, precipitation, water runoff, 
ozone levels, wind speed, seismic events, vegetational patterns, and climate 
dynamics), 
\item medical and physiological time series (recordings of heartbeat, respiration, blood 
pressure, blood flow, nerve spike intervals, human gait, glucose levels, and gene 
expression data),	
\item DNA sequences (they are not actually {\it time} series) , 
\item astrophysical time series (X-ray light sources and sunspot numbers), 
\item technical time series (internet traffic, highway traffic, and neutronic power 
from a reactor), 
\item social time series (finance and economy, language characteristics, fatalities 
in conflicts), 
as well as 
\item physics data (also going beyond {\it time} series), e.~g., surface roughness, 
chaotic spectra of atoms, 
and photon correlation spectroscopy recordings. 
\end{itemize}

If one finds that a complex system is characterised by fractal (or 
multifractal) dynamics with particular scaling exponents, this finding will help 
in obtaining predictions on the future behaviour of the system and on its reaction 
to external perturbations or changes in the boundary conditions.  Phase transitions 
in the regulation behaviour of a complex system are often associated with changes 
in their fractal dynamics, allowing for a detection of such transitions (or the 
corresponding states) by fractal analysis.  One example for a successful application 
of this approach is the human cardiovascular system, where the fractality of 
heartbeat interval time series was shown to reflect certain cardiac impairments as 
well as sleep stages \cite{peng93,bunde00}.  In addition, one can test and iteratively 
improve models of the system until they reproduce the observed scaling behaviour.  
One example for such an approach is climate modelling, where the models were shown 
to need input from volcanos and solar radiation in order to reproduce the long-term 
correlated (fractal) scaling behaviour \cite{vjushin04} previously found in 
observational temperature data \cite{koscielny98}.

Fractal (or multifractal) scaling behaviour certainly cannot be assumed a 
priori, but has to be established.  Hence, there is a need for refined analysis 
techniques, which help to differentiate truly fractal dynamics from fake scaling 
behaviour caused, e.~g., by non-stationarities in the data.  If conventional 
statistical methods are applied for the analysis of time series representing 
the dynamics of a complex system \cite{box94,chatfield03}, there are two major 
problems. (i) The number of data series and their durations (lengths) 
are usually very limited, making it difficult to extract significant information 
on the dynamics of the system in a reliable way.  (ii) If the length of the data 
is extended using computer-based recording techniques or historical (proxy) data, 
non-stationarities in the signals tend to be superimposed upon the intrinsic 
fluctuation properties and measurement noise.  Non-stationarities are 
caused by external or internal effects that lead to either continuous or sudden 
changes in the average values, standard deviations or regulation mechanism.  
They are a major problem for the characterisation of the dynamics, in particular 
for finding the scaling properties of given data.

Following the description of important properties of fractal and multifractal time 
series and the definition of our primary quantities of interest in Section 3, we focus 
on methods for the analysis of self-affine or (mono-)fractal data in Sections 4 and 
5.  While Section 4 describes traditional approaches for the analysis of stationary 
time series, Section 5 discusses more recently established methods applicable for 
non-stationary data.  Section 6 describes techniques for multifractal time series 
analysis.  The consequences of fractal scaling behaviour on the statistics of 
extreme events and their return intervals are presented in Section 7, and a 
few standard models for fractal and multifractal time series are described in 
Section 8 before an outlook on future directions in Section 9.

\section{Fractal and Multifractal Time Series}

\subsection{Fractality, Self-Affinity, and Scaling}

The topic of this article is the fractality (and/or multifractality) of time 
series.  Since fractals and multifractals in general are discussed in many other 
articles of the encyclopedia, the concept is not thoroughly explained here. 
In particular, we refer to the articles ... and ... for the formalism describing
fractal and multifractal structures, respectively.

In a strict sense, most time series are one dimensional, since the 
values of the considered observable are measured in homogeneous time intervals.  
Hence, unless there are missing values, the fractal dimension of the support is
$D(0)=1$.  However, there are rare cases where most of 
the values of a time series are very small or even zero, causing a dimension 
$D(0)<1$ of the support.  In these cases, one has to be very careful in 
selecting appropriate analysis techniques, since many of the methods presented 
in this article are not accurate for such data; the Wavelet Transform Modulus
Maxima technique (see Section 6.2) is the most advanced applicable method.

Even if the fractal dimension of support is one, the information dimension 
$D(1)$ and the correlation dimension $D(2)$ can be studied.  As we will see in 
Section 6.1, $D(2)$ is in fact explicitly related to all exponents studied in 
monofractal time series analysis.  However, usually 
a slightly different approach is employed based on the notion of self-affinity 
instead of (multi-) fractality.  Here, one takes into account that the time axis 
and the axis of the measured values $x(t)$ are not equivalent.  Hence, a rescaling 
of time $t$ by a factor $a$ may require rescaling of the series values $x(t)$ by 
a different factor $a^H$ in order to obtain a statistically similar (i.~e., 
self-similar) picture.  In this case the scaling relation 
\begin{equation} x(t) \to a^H x(a t) \label{self-affine} \end{equation}
holds for an arbitrary factor $a$, describing the data as self-affine (see, e.~g., 
\cite{feder88}).  The {\it Hurst} exponent $H$ (after the water engineer H.E. Hurst 
\cite{hurst51}) characterises the type of self affinity.  Figure {\fig 1}(a) shows
several examples of self-affine time series with different $H$.  
The trace of a random walk (Brownian motion, third line in Fig.~{\fig 1}(a)), 
for example, is characterised by $H=0.5$, 
implying that the position axis must be rescaled by a factor of 2 if the time axis 
is rescaled by a factor of 4.  Note that self-affine series are often denoted as 
fractal even though they are not fractal in the strict sense.  In this article 
the term "fractal" will be used in the more general sense including all data, 
where a Hurst exponent $H$ can be reasonably defined.

The scaling behaviour of self-affine data can also be characterised by looking at
their mean-square displacement.  Since the mean-square displacement of a random 
walker is known to increase linear in time, $\langle x^2(t) \rangle \sim t$, 
deviations from this law will indicate the presence of self-affine scaling.
As we will see in Section 4.4, one can thus retrieve the Hurst (or self-affinity) 
exponent $H$ by studying the scaling behaviour of the mean-square dispalcement, 
or the mean-square fluctuations $\langle x^2(t) \rangle \sim t^{2H}$.

\subsection{Persistence, Long- and Short-term Correlations}

Self-affine data are persistent in the sense that a large value is usually (i.~e., 
with high statistical preference) followed by a large value and a small value is 
followed by a small value.  For the trace of a random walk, persistence on all 
time scales is trivial, since a later position is just a former one plus some
random increment(s).  The persistence holds for all time scales, where the 
self-affinity relation (\ref{self-affine}) holds.  However, the degree of 
persistence can also vary on different time scales.  Weather is a typical example: 
while the weather tomorrow or in one week is probably similar to the weather 
today (due to a stable general weather condition), persistence is much harder 
to be seen on longer time scales. 

Considering the increments $\Delta x_i = x_i - x_{i-1}$ of a self-affine series, 
$(x_i)$, $i=1,\ldots,N$ with $N$ values measured equidistant in time, one finds that 
the $\Delta x_i$ can be either persistent, independent, or anti-persistent.  
Examples for all cases are shown in Fig.~{\fig 1}(b).  In our example of the 
random walk with $H=0.5$ (third line in the figure), the increments (steps) are 
fully independent of each other.  Persistent and anti-persistent increments, 
where a positive increment is likely to be followed by another positive or 
negative increment, respectively, are also leading to persistent integrated series 
$x_i = \sum_{j=1}^i \Delta x_j$.  

For stationary data with constant mean and standard deviation the auto-covariance 
function of the increments,
\begin{equation} C(s) = \big\langle \Delta x_i \, \Delta x_{i+s} \big\rangle = 
{1 \over N-s} \sum_{i=1}^{N-s} \Delta x_i \, \Delta x_{i+s}.
\label{autocorr1}\end{equation}
can be studied to determine the degree of persistence.  If $C(s)$ is divided by the
variance $\langle (\Delta x_i)^2 \rangle$, it becomes the auto-correlation 
function; both are identical if the data are normalised with unit variance.  
If the $\Delta x_i$ are uncorrelated (as for the random walk), $C(s)$ is zero for 
$s>0$.  Short-range correlations of the increments $\Delta x_i$ are usually 
described by $C(s)$ declining exponentially, 
\begin{equation} C(s) \sim \exp(-s/t_\times) \label{src} \end{equation}
with a characteristic decay time $t_\times$.  Such behaviour is typical for 
increments generated by an auto-regressive (AR) process
\begin{equation} \Delta x_i = c \Delta x_{i-1} + \varepsilon_i \label{ARprocess}
\end{equation}
with random uncorrelated offsets $\varepsilon_i$ and $c = \exp(-1/t_\times)$. 
Figure~{\fig 2}(a) shows the auto-correlation function for one configuration 
of an AR process with $t_\times = 48$.

For so-called {\it long-range correlations} $\int_0^\infty C(s)\, ds$ diverges 
in the limit of infinitely long series ($N \to \infty$).  In practice, this means 
that $t_\times$ cannot be defined because it increases with increasing $N$.
For example, $C(s)$ declines as a power-law
\begin{equation} C(s) \propto s^{- \gamma} \label{lrc}\end{equation}
with an exponent $0<\gamma<1$.  Figure~{\fig 2}(b) shows $C(s)$ for one 
configuration with $\gamma = 0.4$.  This type of behaviour can be modelled by 
the Fourier filtering technique (see Section 8.1).  Long-term correlated, i.~e. 
persistent, behaviour of the $\Delta x_i$ leads to self-affine scaling behaviour 
of the $x_i$, characterised by $H=1-\gamma/2$, as will be shown below.

\subsection{Crossovers and Non-stationarities in Time Series}

Short-term correlated increments $\Delta x_i$ characterised by a finite characteristic 
correlation decay time $t_\times$ lead to a crossover in the scaling behaviour of the 
integrated series $x_i = \sum_{j=1}^i \Delta x_j$, see Fig.~{\fig 2}(a) for an example. 
Since the position of the crossover might be numerically different from $t_\times$,
we denote it by $s_\times$ here.  Time series with a crossover are not self-affine 
and there is no unique Hurst exponent $H$ characterising them.  While $H>0.5$ is 
observed on small time scales (indicating correlations in the increments), the 
asymptotic behaviour (for large time scales $s \gg t_\times$ and $\gg s_\times$) 
is always characterised by $H=0.5$, since all correlations have decayed.  
Many natural recordings are characterised by pronounced short-term correlations 
in addition to scaling long-term correlations.  For example, there are short-term 
correlations due to particular general weather situations in temperature data and 
due to respirational effects in heartbeat data.  Crossovers in the scaling 
behaviour of complex time series can also be caused by different regulation 
mechanisms on fast and slow time scales.  Fluctuations of river runoff, for 
example, show different scaling behaviour on time scales below and above 
approximately one year.  

Non-stationarities can also cause crossovers in the scaling behaviour of data
if they are not properly taken into account.  In the most strict sense, 
non-stationarities are variations in the mean or the standard deviation of the 
data (violating weak stationarity) or the distribution of the data values 
(violating strong stationarity).  Non-stationarities like monotonous, periodic 
or step-like trends are often caused by external effects, e.~g., by the 
greenhouse warming and seasonal variations for temperature records, different 
levels of activity in long-term physiological data, or unstable light sources 
in photon correlation spectroscopy.  Another example for non-stationary data is a
record consisting of segments with strong fluctuations alternating with segments 
with weak fluctuations.  Such behaviour will cause a crossover in scaling at the 
time scale corresponding to the typical duration of the homogeneous segments.  
Different mechanisms of regulation during different time segments -- like, e.~g., 
different heartbeat regulation during different sleep stages at night -- can 
also cause crossovers; they are regarded as non-stationarities here, too.  
Hence, if crossovers in the scaling behaviour of data are observed, more 
detailed studies are needed to find out the cause of the crossovers.  One can 
try to obtain homogenous data by splitting the original series and employing 
methods that are at least insensitive to monotonous (polynomially shaped) trends.  

To characterise a complex system based on time series, trends and fluctuations are 
usually studied separately (see, e.~g., \cite{schmitt06} for a discussion).  Strong 
trends in data can lead to a false detection of long-range statistical persistence 
if only one (non-detrending) method is used or if the results are not carefully 
interpreted.  Using several advanced techniques of scaling time series analysis 
(as described in Chapter 5) crossovers due to trends can be distinguished from 
crossovers due to different regulation mechanisms on fast and slow time scales.  
The techniques can thus assists in gaining insight into the scaling behaviour of 
the natural variability as well as into the kind of trends of the considered time 
series.  

It has to be stressed that crossovers in scaling behaviour must not be confused 
with multifractality.  Even though several scaling exponents are needed, they 
are not applicable for the same regime (i.~e., the same range of time scales).  
Real multifractality, on the other hand, is characterised by different scaling 
behaviour of different moments over the full range of time scales (see next 
section).

\subsection{Multifractal Time Series}

Many records do not exhibit a simple monofractal scaling behaviour, which can be 
accounted for by a single scaling exponent.  As discussed in the previous section,
there might exist crossover (time-) scales $s_\times$ separating regimes with 
different scaling exponents.  In other cases, the scaling behaviour is more 
complicated, and different scaling exponents are required for different parts of 
the series.  In even more complicated cases, such different scaling behaviour can 
be observed for many interwoven fractal subsets of the time series.  In this case 
a multitude of scaling exponents is required for a full description of the scaling 
behaviour in the same range of time scales, and a multifractal analysis must be 
applied.

Two general types of multifractality in time series can be distinguished: 
(i) Multifractality due to a broad probability distribution (density function) 
for the values of the time series, e.~g. a Levy distribution.  In this case the 
multifractality cannot be removed by shuffling the series. (ii) Multifractality 
due to different long-term correlations of the small and large fluctuations.  
In this case the probability density function of the values can be a regular 
distribution with finite moments, e.~g., a Gaussian distribution.  The corresponding 
shuffled series will exhibit non-multifractal scaling, since all long-range 
correlations are destroyed by the shuffling procedure.  Randomly shuffling the 
order of the values in the time series is the easiest way of generating surrogate
data; however, there are more advanced alternatives (see Chapter 8).  If both 
kinds of multifractality are present, the shuffled series will show weaker 
multifractality than the original series.

A multifractal analysis of time series will also reveal higher order correlations.
Multifractal scaling can be observed if, e.~g., three or four-point correlations
scale differently from the standard two-point correlations studied by classical
autocorrelation analysis (Eq.~(\ref{autocorr1})).  In addition, multifractal scaling
is observed if the scaling behaviour of small and large fluctuations is different.
For example, extreme events might be more or less correlated than typical events.

\section{Methods for Stationary Fractal Time Series Analysis}

In this chapter we describe four traditional approaches for the fractal analysis 
of stationary time series, see \cite{Taqqu95,delignieresa06,Mielniczuk07} for
comparative studies.  The main focus is on the determination of the scaling
exponents $H$ or $\gamma$, defined in Eqs.~(\ref{self-affine}) and (\ref{lrc}),
respectively, and linked by $H=1-\gamma/2$ in long-term persistent data.  Methods
taking non-stationarities into account will be discussed in the next chapter.

\subsection{Autocorrelation Function Analysis}

We consider a record $(x_i)$ of $i=1,\ldots,N$ equidistant measurements.  In 
most applications, the index $i$ will correspond to the time of the measurements.
We are interested in the correlation of the values $x_i$ and $x_{i+s}$ for 
different time lags, i.~e. correlations over different time scales $s$.  In order 
to remove a constant offset in the data, the mean $\langle x \rangle = {1 \over N} 
\sum_{i=1}^{N} x_i$ is usually subtracted, $\tilde {x}_i \equiv x_i - \langle x 
\rangle$.  Alternatively, the correlation properties of increments $\tilde x_i =
\Delta x_i = x_i - x_{i-1}$ of the original series can be studied (see also Section 
3.2).  Quantitatively, correlations between $\tilde x$-values separated by $s$ steps 
are defined by the (auto-) covariance function $C(s) = \langle \tilde{x}_i \,
\tilde{x}_{i+s}\rangle$ or the (auto-) correlation function $C(s) / \langle 
\tilde{x}_i^2 \rangle$, see also Eq.~(\ref{autocorr1}).

As already mentioned in Section 3.2, the $\tilde x_i$ are short-term correlated if 
$C(s)$ declines exponentially, $C(s) \sim \exp(-s/t_\times)$, and long-term 
correlated if $C(s)$ declines as a power-law $C(s) \propto s^{-\gamma}$ with a 
correlation exponent $0 < \gamma < 1$ (see Eqs.~(\ref{src}) and (\ref{lrc}),
respectively).  As illustrated by the two examples shown in Fig.~{\fig 2}, a 
direct calculation of $C(s)$ is usually not appropriate due to noise superimposed 
on the data $\tilde x_i$ and due to underlying non-stationarities of 
unknown origin.  Non-stationarities make the definition of $C(s)$ problematic, 
because the average $\langle x \rangle$ is not well-defined.  Furthermore, $C(s)$ 
strongly fluctuates around zero on large scales $s$ (see Fig.~2(b)), making it 
impossible to find the correct correlation exponent $\gamma$.  Thus, one has to 
determine the value of $\gamma$ indirectly.

\subsection{Spectral Analysis}

If the time series is stationary, we can apply standard spectral analysis techniques 
(Fourier transform) and calculate the power spectrum $S(f)$ of the time series 
$(\tilde x_i)$ as a function of the frequency $f$ to determine self-affine scaling 
behaviour \cite{Hunt51}.  For long-term correlated data characterised by the 
correlation exponent $\gamma$, we have 
\begin{equation} S(f) \sim f^{-\beta} \quad {\rm with} \quad \beta= 1-\gamma.
\label{spectr}\end{equation}
The spectral exponent $\beta$ and the correlation exponent $\gamma$ can thus be 
obtained by fitting a power-law to a double logarithmic plot of the power spectrum 
$S(f)$.  An example is shown in Fig.~{\fig 3}.  The relation (\ref{spectr}) can be 
derived from the Wiener-Khinchin theorem (see, e.~g., \cite{Rangarajan00}).  If, 
instead of $\tilde x_i = \Delta x_i$ the integrated runoff time series is Fourier 
transformed, i.~e., $\tilde x_i = x_i  \sum_{j=1}^i \Delta x_j$, the resulting 
power spectrum scales as $S(f) \sim f^{-2-\beta}$.  

Spectral analysis, however, does not yield more reliable results than auto-correlation
analysis unless a logarithmic binning procedure is applied to the double logarithmic
plot of $S(f)$ \cite{Taqqu95}, see also Fig.~{\fig 3}.  I.~e., the average of 
$\log S(f)$ is calculated in successive, logarithmically wide bands from $a^n f_0$ 
to $a^{n+1} f_0$, where $f_0$ is the minimum frequency, $a>1$ is a factor (e.~g., 
$a=1.1$), and the index $n$ is counting the bins.  Spectral analysis also requires 
stationarity of the data.

\subsection{Hurst's Rescaled-Range Analysis}

The first method for the analysis of long-term persistence in time series based on 
random walk theory has been proposed by the water construction engineer Harold Edwin 
Hurst (1880-1978), who developed it while working in Egypt.  His so-called rescaled 
range analysis ($R/S$ analysis) \cite{mandel68,mandel69,hurst51,hurst65,feder88} 
begins with splitting of the time series $(\tilde x_i)$ into non-overlapping 
segments $\nu$ of size (time scale) $s$ (first step), yielding $N_s= {\rm int}(N/s)$ 
segments altogether.  In the second step, the {\it profile} (integrated data) is 
calculated in each segment $\nu=0,\ldots,N_s-1$,
\begin{equation} Y_\nu(j) = \sum_{i=1}^j \left( \tilde x_{\nu s + i} - \langle 
\tilde x_{\nu s + i} \rangle_s \right) = \sum_{i=1}^j \tilde x_{\nu s + i} -
{j \over s} \sum_{i=1}^s \tilde x_{\nu s + i}. \label{Yrs}\end{equation}
By the subtraction of the local averages, piecewise constant trends in the data 
are eliminated.  In the third step, the differences between minimum and maximum 
value ({\it ranges}) $R_\nu(s)$ and the standard deviations $S_\nu(s)$ in each 
segment are calculated,
\begin{equation} R_\nu(s) = {\rm max}_{j=1}^s Y_\nu(j) - {\rm min}_{j=1}^s Y_\nu(j), 
\quad S_\nu(s) = \sqrt{ {1 \over s} \sum_{j=1}^s Y^2_\nu(j)}. 
\label{rs2} 
\end{equation}
Finally, the rescaled range is averaged over all segments to obtain the fluctuation
function $F(s)$,
\begin{equation} F_{RS}(s) = {1 \over N_s} \sum_{\nu=0}^{N_s-1} {R_\nu(s) \over 
S_\nu(s)} \sim s^H \qquad {\rm for} \quad s \gg 1,
\label{rs3}\end{equation}
where $H$ is the Hurst exponent already introduced in Eq.~(\ref{self-affine}). 
One can show \cite{mandel68,Hunt51} that $H$ is related to $\beta$ and $\gamma$ 
by $2H \approx 1+\beta = 2-\gamma$ (see also Eqs.~(\ref{spectr}) and (\ref{alpha})).
Note that $0<\gamma<1$, so that the right part of the equation does not hold unless 
$0.5<H<1$.  The relationship does {\it not} hold in general for multifractal data.  
Note also that $H$ actually characterises the self-affinity of the profile function 
(\ref{Yrs}), while $\beta$ and $\gamma$ refer to the original data.

The values of $H$, that can be obtained by Hurst's rescaled range 
analysis, are limited to $0<H<2$, and significant inaccuracies are to be expected 
close to the bounds.  Since $H$ can be increased or decreased by 1 if the data is 
integrated ($\tilde x_j \to \sum_{i=1}^j \tilde x_i$) or differentiated ($\tilde x_i
\to \tilde x_i - \tilde x_{i-1}$), respectively, one can always find a way to 
calculate $H$ by rescaled range analysis provided the data is stationary.  While 
values $H<1/2$ indicate long-term anti-correlated behaviour of the data $\tilde x_i$, 
$H>1/2$ indicates long-term positively correlated behaviour.  For power-law 
correlations decaying faster than $1/s$, we have $H=1/2$ for large $s$ values, like 
for uncorrelated data. 

Compared with spectral analysis, Hurst's rescaled range analysis yields smoother 
curves with less effort (no binning procedure is necessary) and works also for data 
with piecewise constant trends.

\subsection{Fluctuation Analysis (FA)}

The standard fluctuation analysis (FA) \cite{bunde94,peng92} is also based on 
random walk theory.  For a time series $(\tilde x_i)$, $i=1,\ldots,N$, with zero 
mean, we consider the global profile, i.~e., the cumulative sum (cf. Eq.~(\ref{Yrs}))
\begin{equation} Y(j) = \sum_{i=1}^j \tilde x_i, \quad j=0,1,2,\ldots,N,
\label{profile}\end{equation}
and study how the fluctuations of the profile, in a given time window of size $s$, 
increase with $s$.  The procedure is illustrated in Fig.~{\fig 4} for two values
of $s$.  We can consider the profile $Y(j)$ as the position of a random 
walker on a linear chain after $j$ steps. The random walker starts at the origin 
and performs, in the $i$th step, a jump of length $\tilde x_i$ to the right, if  
$\tilde x_i$ is positive, and to the left, if $\tilde x_i$ is negative.

To find how the square-fluctuations of the profile scale with $s$, we first divide 
each record of $N$ elements into $N_s= {\rm int}(N/s)$ non-overlapping segments 
of size $s$ starting from the beginning (see Fig.~{\fig 4}) and another $N_s$ 
non-overlapping segments of size $s$ starting from the end of the considered series.  
This way neither data at the end nor at the beginning of the record is neglected.  
Then we determine the fluctuations in each segment $\nu$.

In the standard FA, we obtain the fluctuations just from the 
values of the profile at both endpoints of each segment $\nu=0,\ldots,N_s-1$, 
\begin{equation} F_{\rm FA}^2(\nu,s) = [Y(\nu s) - Y((\nu+1) s)]^2, 
\label{fa2}\end{equation}
(see Fig.~{\fig 4}) and analogous for $\nu=N_s,\ldots,2N_s-1$, 
\begin{equation} F_{\rm FA}^2(\nu,s) = [Y(N-(\nu-N_s) s) - Y(N-(\nu+1-N_s) s)]^2. 
\label{fa3}\end{equation}
Then we average $F_{\rm FA}^2(\nu,s)$ over all subsequences to obtain the mean 
fluctuation $F_2(s)$,
\begin{equation} F_2(s) = \left[ {1 \over 2 N_s} \sum_{\nu=0}^{2 N_s-1} 
F_{\rm FA}^2(\nu,s)\right]^{1/2} \sim s^\alpha. \label{Fs2}\end{equation}
By definition, $F_2(s)$ can be viewed as the root-mean-square displacement of the 
random walker on the chain, after $s$ steps (the reason for the index 2 will 
become clear later).  For uncorrelated $x_i$ values, we obtain Fick's diffusion 
law $F_2(s) \sim s^{1/2}$.  For the relevant case of long-term correlations, 
where $C(s)$ follows the power-law behaviour of Eq.~(\ref{lrc}), $F_2(s)$ 
increases by a power law,
\begin{equation} F_2(s) \sim s^\alpha \quad {\rm with} \quad \alpha \approx H, 
\label{alpha}\end{equation}
where the fluctuation exponent $\alpha$ is identical with the Hurst exponent $H$
for mono-fractal data and related to $\gamma$ and $\beta$ by 
\begin{equation} 2\alpha = 1+\beta = 2-\gamma.
\label{exponents}\end{equation}
The typical behaviour of $F_2(s)$ for short-term correlated and long-term correlated 
data is illustrated in Fig.~{\fig 2}.  The relation (\ref{exponents}) can be derived 
straightforwardly by inserting Eqs.~(\ref{profile}), (\ref{autocorr1}), and (\ref{lrc}) 
into Eq.~(\ref{fa2}) and separating sums over products $\tilde x_i \tilde x_j$ with 
identical and different $i$ and $j$, respectively.

The range of the $\alpha$ values that can be studied by standard FA is limited 
to $0 < \alpha < 1$, again with significant inaccuracies close to the bounds.  
Regarding integration or differentiation of the data, the same rules apply as 
listed for $H$ in the previous section.  The results of FA become statistically 
unreliable for scales $s$ larger than one tenth of the length of the data, i.~e. 
the analysis should be limited by $s<N/10$.

\section{Methods for Non-Stationary Fractal Time-Series Analysis}

\subsection{Wavelet Analysis}

The origins of wavelet analysis come from signal theory, where frequency 
decompositions of time series were studied \cite{Goupillaud,Daubechies}. 
Like the Fourier transform, the wavelet transform of a signal $x(t)$ is a 
convolution integral to be replace by a summation in case of a discrete
time series $(\tilde x_i), i=1,\ldots,N$,
\begin{equation} L_\psi(\tau,s) = {1 \over s} \int_{-\infty}^\infty x(t)
\psi[(t-\tau)/s] \, dt = {1 \over s} \sum_{i=1}^N  \tilde x_i \, \psi[(i-\tau)/s]. 
\label{wt}\end{equation}
Here, $\psi(t)$ is a so-called mother wavelet, from which all daughter
wavelets $\psi_{\tau,s}(t) = \psi((t-\tau)/s$ evolve by shifting and 
stretching of the time axis.  The wavelet coefficients $L_\psi(\tau,s)$ thus 
depend on both, time position $\tau$ and scale $s$.  Hence, the local 
frequency decomposition of the signal is described with a time resolution 
appropriate for the considered frequency $f=1/s$ (i.~e., inverse time scale).

All wavelets $\psi(t)$ must have zero mean.  They are often chosen to be 
orthogonal to polynomial trends, so that the analysis method becomes 
insensitive to possible trends in the data.  Simple examples are derivatives 
of a Gaussian, $\psi_{\rm Gauss}^{(n)}(t)={d^n \over dt^n} \exp(-x^2/2)$, 
like the Mexican hat wavelet $-\psi_{\rm Gauss}^{(2)}$ and the Haar wavelet, 
$\psi_{\rm Haar}^{(0)}(t) = +1$ if $0\le t<1$, $-1$ if $1\le t<2$, and 0 
otherwise.  It is straightforward to construct Haar wavelet that are 
orthogonal to linear, quadratic and cubic trends, e.~g., 
$\psi_{\rm Haar}^{(1)}(t) = 1$ for $t \in [0,1) \cup [2,3)$, $-2$ for $t 
\in [1,2)$, and 0 otherwise, or $\psi_{\rm Haar}^{(2)}(t) = 1$ for $t \in 
[0,1)$, $-3$ for $t \in [1,2)$, $+3$ for $t \in [2,3)$, $-1$ for $t \in 
[3,4)$, and 0 otherwise.

\subsection{Discrete Wavelet Transform (WT) Approach}

A detrending fractal analysis of time series can be easily implemented by
considering Haar wavelet coefficients of the profile $Y(j)$, 
Eq.~(\ref{profile}) \cite{physa97,koscielny98}.  In this case the convolution 
(\ref{wt}) corresponds to the addition and subtraction of mean values of 
$Y(j)$ within segments of size $s$.  Hence, defining $\bar Y_\nu(s) = 
{1 \over s} \sum_{j=1}^s Y(\nu s + j)$, the coefficients can be written as
\begin{equation} F_{\rm WT0}(\nu,s) \equiv L_{\psi_{\rm Haar}^{(0)}}(\nu s,s) 
= \bar Y_\nu(s) - \bar Y_{\nu+1}(s), \label{wt0} \end{equation}
\begin{equation} F_{\rm WT1}(\nu,s) \equiv L_{\psi_{\rm Haar}^{(1)}}(\nu s,s) 
= \bar Y_\nu(s) - 2 \bar Y_{\nu+1}(s) + \bar Y_{\nu+2}(s), \quad 
{\rm and} \label{wt1} \end{equation}
\begin{equation} F_{\rm WT2}(\nu,s) \equiv L_{\psi_{\rm Haar}^{(2)}}(\nu s,s) 
= \bar Y_\nu(s) - 3 \bar Y_{\nu+1}(s) + 3 \bar Y_{\nu+2}(s) - \bar 
Y_{\nu+3}(s) \label{wt2} \end{equation}
for constant, linear and quadratic detrending, respectively.  The 
generalization for higher orders of detrending is obvious.  The resulting 
mean-square fluctuations $F^2_{{\rm WT}n}(\nu,s)$ are averaged over all $\nu$ 
to obtain the mean fluctuation $F_2(s)$, see Eq.~(\ref{Fs2}).  Figure {\fig 5} 
shows typical results for WT analysis of long-term correlated, short-term 
correlated and uncorrelated data.

Regarding trend-elimination, wavelet transform WT0 corresponds to standard 
FA (see Section 4.4), and only constant trends in the profile are eliminated.  
WT1 is similar to Hurst's rescaled range analysis (see Section 4.3): linear 
trends in the profile and constant trends in the data are eliminated, and the 
range of the fluctuation exponent $\alpha \approx H$ is up to 2.  In general, 
WT$n$ determines the fluctuations from the $n$th derivative, this way 
eliminating trends described by $(n-1)$st-order polynomials in the data.  
The results become statistically unreliable for scales $s$ larger than one 
tenth of the length of the data, just as for FA.

\subsection{Detrended Fluctuation Analysis (DFA)}

In the last 13 years {\it Detrended Fluctuation Analysis} (DFA), originally 
introduced by Peng et al. \cite{peng94}, has been established as an important 
method to reliably detect long-range (auto-) correlations in non-stationary 
time series.  The method is based on random walk theory and basically represents 
a linear detrending version of FA (see Section 4.4).  DFA was later generalised 
for higher order detrending \cite{bunde00}, separate analysis of sign and 
magnitude series \cite{PRL01} (see Section 5.5), multifractal analysis 
\cite{kantelhardt02} (see Section 6.3), and data with more than one dimension 
\cite{gu06}.  Its features have been studied in many articles 
\cite{kantelhardt01,hu01,chen02,chen03,grau06,nagarajan06}.  In addition, 
several comparisons of DFA with other methods for stationary and non-stationary 
time-series analysis have been published, see, e.~g., 
\cite{Taqqu95,Mielniczuk07,heneghan00,Weron02} and in particular 
\cite{delignieresa06}, where DFA is compared with many other established methods
for short data sets, and \cite{amir07}, where it is compared with recently 
suggested improved methods.  Altogether, there are about 450-500 papers applying 
DFA (till April 2008).  In most cases positive auto-correlations were reported 
leaving only a few exceptions with anti-correlations, see, e.~g., 
\cite{bahar01,ekg-form,santhanam06}. 

Like in the FA method, one first calculates the global profile according to 
Eq.~(\ref{profile}) and divides the profile into $N_s = {\rm int}(N/s)$ non-overlapping 
segments of size $s$ starting from the beginning and another $N_s$ segments 
starting from the end of the considered series.  DFA explicitly deals with 
monotonous trends in a detrending procedure.  This is done by estimating a 
polynomial trend $y_{\nu,s}^m(j)$ within each segment $\nu$ by least-square 
fitting and subtracting this trend from the original profile (`detrending'),
\begin{equation} \tilde{Y}_s(j)=Y(j)-y_{\nu,s}^{m}(j). \label{detrendedfunction} 
\end{equation}
The degree of the polynomial can be varied in order to eliminate constant ($m=0$),
linear $(m=1)$, quadratic $(m=2)$ or higher order trends of the profile function 
\cite{bunde00}.  Conventionally the DFA is named after the order of the fitting 
polynomial (DFA0, DFA1, DFA2, ...).   In DFA$m$, trends of order $m$ in the 
profile $Y(j)$ and of order $m-1$ in the original record $\tilde x_i$ are 
eliminated.  The variance of the detrended profile $\tilde{Y}_s(j)$ in each 
segment $\nu$ yields the mean-square fluctuations,
\begin{equation} F^2_{{\rm DFA}m}(\nu,s)=\frac{1}{s} \sum_{j=1}^s 
\tilde{Y}_s^2(j). \label{dfap} \end{equation}
As for FA and discrete wavelet analysis, the $F^2_{{\rm DFA}m}(\nu,s)$ are 
averaged over all segments $\nu$ to obtain the mean fluctuations $F_2(s)$, see 
Eq.~(\ref{alpha}).  Calculating $F_2(s)$ for many $s$, the fluctuation scaling 
exponent $\alpha$ can be determined just as with FA.  Figure {\fig 6} 
shows typical results for DFA of the same long-term correlated, short-term 
correlated and uncorrelated data studied already in Fig.~{\fig 5}.

We note that in studies that include averaging over many records (or one record 
cut into many separate pieces by the elimination of some unreliable intermediate 
data points) the averaging procedure (\ref{Fs2}) must be performed for all data.  
Taking the square root is always the final step after all averaging is finished.  
It is not appropriate to calculate $F_2(s)$ for parts of the data and then average 
the $F_2(s)$ values, since such a procedure will bias the results towards smaller 
scaling exponents on large time scales.

If $F_2(s)$ increases for increasing $s$ by $F_2(s)\sim s^\alpha$ with $0.5<\alpha
<1$, one finds that the scaling exponent $\alpha \approx H$ is related to the 
correlation exponent $\gamma$ by $\alpha=1-\gamma/2$ (see Eq.~(\ref{exponents})). 
A value of $\alpha=0.5$ thus indicates that there are no (or only short-range) 
correlations. If $\alpha>0.5$ for all scales $s$, the data are long-term 
correlated.  The higher $\alpha$, the stronger the correlations in the signal 
are.  $\alpha>1$ indicates a non-stationary local average of the data; in this 
case, FA fails and yields only $\alpha=1$.  The case $\alpha<0.5$ corresponds to 
long-term anti-correlations, meaning that large values are most likely to be 
followed by small values and vice versa.  $\alpha$ values below 0 er not possible.
Since the maximum value for $\alpha$ in DFA$m$ is $m+1$, higher detrending orders 
should be used for very non-stationary data with large $\alpha$.  Like in FA and 
Hurst's analysis, $\alpha$ will decrease or increase by one upon additional 
differentiation or integration of the data, respectively.

Small deviations from the scaling law (\ref{alpha}), i.~e. deviations from a 
straight line in a double logarithmic plot, occur for small scales $s$, in
particular for DFA$m$ with large detrending order $m$.  These deviations are 
intrinsic to the usual DFA method, since the scaling behaviour is only approached 
asymptotically.  The deviations limit the capability of DFA to determine the 
correct correlation behaviour in very short records and in the regime of small 
$s$.  DFA6, e.~g., is only defined for $s \ge 8$, and significant deviations 
from the scaling law $F_2(s) \sim s^\alpha$ occur even up to $s \approx 30$.  
They will lead to an over-estimation of the fluctuation exponent $\alpha$, 
if the regime of small $s$ is used in a fitting procedure.  An approach for 
correction of this systematic artefact in DFA is described in 
\cite{kantelhardt01}.

The number of independent segments of length $s$ is larger in DFA than in WT, 
and the fluctuations in FA are larger than in DFA.  Hence, the analysis has to
be based on $s$ values lower than $s_{\rm max} = N/4$ for DFA compared with 
$s_{\rm max} = N/10$ for FA and WT.  The accuracy of scaling exponents $\alpha$
determined by DFA was recently studied as a function of the length $N$ of the 
data \cite{amir07} (fitting range $s \in [10,N/2]$ was used).  The results show 
that statistical standard errors of $\alpha$ (one standard deviation) are 
approximately 0.1 for $N=500$, 0.05 for $N=$ 3 000, and reach $0.03$ for $N=$ 
10 000.  Findings of long-term correlations with $\alpha=0.6$ in data with only 
500 points are thus not significant; and $\alpha$ should be at least 0.55 even
for data of 10 000 points.

A generalization of DFA for two-dimensional data (or even higher dimensions $d$)
was recently suggested \cite{gu06}.  The generalization works well when tested 
with synthetic surfaces including fractional Brownian surfaces and multifractal 
surfaces. The two-dimensional MFDFA is also adopted to analyse two images from 
nature and experiment, and nice scaling laws are unravelled. In the 2d procedure, 
a double cumulative sum (profile) is calculated by summing over both directional 
indices analogous with Eq.~(\ref{profile}), $Y(k,l)=\sum_{i=1}^k \sum_{j=1}^l 
\tilde x_{i,j}$.  This surface is partitioned into squares of size $s\times s$ 
with indices $\nu$ and $\mu$, in which polynomials like $y_{\nu,\mu,s}^{2}(i,j) = 
ai^2 + bj^22 + cij + di + ej + f$ are fitted. The fluctuation function $F_2(s)$ is again 
obtained by calculating the variance of the profile from the fits.

\subsection{Detection of Trends and Crossovers with DFA}

Frequently, the correlations of recorded data do not follow the same scaling
law for all time scales $s$, but one or sometimes even more crossovers between
different scaling regimes are observed (see Section 3.3).  Time series with a 
well-defined crossover at $s_\times$ and vanishing correlations above $s_\times$ 
are most easily generated by Fourier filtering (see Section 8.1).  The power 
spectrum $S(f)$ of an uncorrelated random series is multiplied by 
$(f/f_\times)^{-\beta}$ with $\beta = 2 \alpha -1$ for frequencies $f > 
f_\times = 1/s_\times$ only.  The series obtained by inverse Fourier transform 
of this modified power spectrum exhibits power-law correlations on time scales 
$s < s_\times$ only, while the behaviour becomes uncorrelated on larger time 
scales $s > s_\times$.  

The crossover from $F_2(s) \sim s^\alpha$ to $F_2(s) \sim s^{1/2}$ is clearly 
visible in double logarithmic plots of the DFA fluctuation function for such
short-term correlated data.  However, it occurs at times $s_\times^{(m)}$ 
that are different from the original $s_\times$ used for the generation of 
the data and that depend on the detrending order $m$.  This systematic 
deviation is most significant in the DFA$m$ with higher $m$.  
Extensive numerical simulations (see Fig.~3 in \cite{kantelhardt01}) show 
that the ratios of $s_\times^{(m)}/s_\times$ are 1.6, 2.6, 3.6, 4.5, and 5.4 
for DFA1, DFA2, \dots, DFA5, with an error bar of approximately 0.1.  Note,
however, that the precise value of this ratio will depend on the method used
for fitting the crossover times $s_\times^{(m)}$ (and the method used for 
generating the data if generated data is analysed).  If results for different
orders of DFA shall be compared, an 
observed crossover $s_\times^{(m)}$ can be systematically corrected 
dividing by the ratio for the corresponding DFA$m$.  If several orders of 
DFA are used in the procedure, several estimates for the real $s_\times$ 
will be obtained, which can be checked for consistency or used for an error 
approximation.  A real crossover can thus be well distinguished from the 
effects of non-stationarities in the data, which lead to a different dependence 
of an apparent crossover on $m$.  

The procedure is also required if the characteristic time scale of short-term 
correlations shall be studied with DFA.  If consistent (corrected) $s_\times$ 
values are obtained based on DFA$m$ with different $m$, the existence of a 
real characteristic correlation time scale is positively confirmed.  Note that 
lower detrending orders are advantageous in this case, since the observed 
crossover time scale $s_\times^{(m)}$ might become quite large and nearly 
reach one forth of the total series length ($N/4$), where the results become 
statistically inaccurate.

We would like to note that studies showing scaling long-term correlations 
should not be based on DFA or variants of this method alone in most 
applications. In particular, if it is not clear whether a given time series 
is indeed long-term correlated or just short-term correlated with a fairly 
large crossover time scale, results of DFA should be compared with other 
methods.  For example, one can employ wavelet methods (see, e.~g., Section 
5.2). Another option is to remove short-term correlations by considering 
averaged series for comparison.  For a time series with daily observations 
and possible short-term correlations up to two years, for example, one might 
consider the series of two-year averages and apply DFA together with FA, 
binned power spectra analysis, and/or wavelet analysis.  Only if these methods 
still indicate long-term correlations, one can be sure that the data are indeed 
long-term correlated.

As discussed in Section 3.3, records from real measurements are often affected 
by non-stationarities, and in particular by trends.  They have to be well 
distinguished from the intrinsic fluctuations of the system.  To investigate 
the effect of trends on the DFA$m$ fluctuation functions, one can generate 
artificial series $(\tilde x_i)$ with smooth monotonous trends by adding 
polynomials of different power $p$ to the original record $(x_i)$, 
\begin{equation}
\tilde x_i = x_i + A x^p \qquad {\rm with} \quad x = i/N. \label{trend}
\end{equation}
For the DFA$m$, such trends in the data can lead to an artificial crossover 
in the scaling behaviour of $F_2(s)$, i.~e., the slope $\alpha$ is strongly 
increased for large time scales $s$.  The position of this artificial 
crossover depends on the strength $A$ and the power $p$ of the trend.  
Evidently, no artificial crossover is observed, if the detrending order $m$ 
is larger than $p$ and $p$ is integer.  The order $p$ of the trends in the 
data can be determined easily by applying the different DFA$m$.  If $p$ is 
larger than $m$ or $p$ is not an integer, an artificial crossover is observed, 
the slope $\alpha_{\rm trend}$ in the large $s$ regime strongly depends on $m$, 
and the position of the artificial crossover also depends strongly on $m$. The 
artificial crossover can thus be clearly distinguished from real crossovers in 
the correlation behaviour, which result in identical slopes $\alpha$ and rather 
similar crossover positions for all detrending orders $m$.  For more extensive 
studies of trends with non-integer powers we refer to \cite{kantelhardt01,hu01}.  
The effects of periodic trends are also studied in \cite{kantelhardt01}.

If the functional form of the trend in given data is not known a priori, the fluctuation 
function $F_2(s)$ should be calculated for several orders $m$ of the fitting 
polynomial.  If $m$ is too low, $F_2(s)$ will show a pronounced crossover to a 
regime with larger slope for large scales $s$ \cite{kantelhardt01,hu01}.  The 
maximum slope of $\log F_2(s)$ versus $\log s$ is $m+1$.  The crossover will 
move to larger scales $s$ or disappear when $m$ is increased, unless it is a 
real crossover not due to trends.  Hence, one can find $m$ such that detrending 
is sufficient.  However, $m$ should not be larger than necessary, because 
shifts of the observed crossover time scales and deviations on short scales
$s$ increase with increasing $m$.

\subsection{Sign and Magnitude (Volatility) DFA}

To study the origin of long-term fractal correlations in a time series, the 
series can be split into two parts, which are analysed separately.  It is 
particularly useful to split the series of increments, $\Delta x_i =
x_i - x_{i-1}$, $i=1, \ldots, N$, into a series of signs $\tilde x_i = s_i
= {\rm sign} \Delta x_i$ and a series of magnitudes $\tilde x_i = m_i = 
\vert \Delta x_i \vert$ \cite{PRL01,ashkenazy03,kalisky05}.  There is an extensive
interest in the magnitude time series in economics \cite{stanley,bouchaud}.  
These data, usually called volatility, represents the absolute variations in 
stock (or commodity) prices and are used as a measure quantifying the risk of
investments.  While the actual prices are only short-term correlated, long-term 
correlations have been observed in volatility series \cite{stanley,bouchaud}. 

Time series having identical distributions 
and long-range correlation properties can exhibit quite different temporal 
organizations of the magnitude and sign sub-series.  The DFA method can be 
applied independently to both of these series.  Since in particular the signs 
are often rather strongly anti-correlated and DFA will give incorrect results 
if $\alpha$ is too close to zero, one often studies integrated sign and magnitude 
series. As mentioned above, integration $\tilde x_i \to \sum_{j=1}^i \tilde x_j$
increases $\alpha$ by one.

Most published results report short-term anti-correlations and no long-term 
correlations in the sign series, i.~e., $\alpha_{\rm sign} < 1/2$ for the 
non-integrated signs $s_i$ (or $\alpha_{\rm sign} < 3/2$ for the integrated 
signs) on low time scales and $\alpha_{\rm sign} \to 1/2$ asymptotically for 
large $s$.  The magnitude series, on the other hand, are usually either 
uncorrelated $\alpha_{\rm magn} = 1/2$ (or $3/2$) or positively long-term 
correlated $\alpha_{\rm magn} > 1/2$ (or $3/2$).  It has been suggested that 
findings of $\alpha_{\rm magn} > 1/2$ are related with nonlinear properties 
of the data and in particular multifractality \cite{PRL01,ashkenazy03,kalisky05}, 
if $\alpha<1.5$ in standard DFA.  Specifically, the results suggest that the 
correlation exponent of the magnitude series is a monotonically increasing 
function of the multifractal spectrum (i.~e., the singularity spectrum) width 
of the original series (see Section 6.1). On the other hand, the sign series 
mainly relates to linear properties of the original series. At small time scales 
$s<16$ the standard $\alpha$ is approximately the average of  $\alpha_{\rm sign}$ 
and $\alpha_{\rm magn}$, if integrated sign and magnitude series are analysed.
For $\alpha>1.5$ in the original series, the integrated magnitude and sign series 
have approximately the same two-point scaling exponents \cite{ashkenazy03}.
An analytical treatment is presented in \cite{kalisky05}.

\subsection{Further Detrending Approaches}

A possible drawback of the DFA method is the occurrence of abrupt jumps in the 
detrended profile $\tilde{Y}_s(j)$ (Eq.~(\ref{detrendedfunction})) at the boundaries 
between the segments, since the fitting polynomials in neighbouring segments are not 
related. A possible way to avoid these jumps would be the calculation of $F_2(s)$ 
based on polynomial fits in overlapping windows. However, this is rather time 
consuming due to the polynomial fit in each segment and is consequently not done 
in most applications. To overcome the problem of jumps several modifications and
extensions of the FA and DFA methods have been suggested in the last years.  
These methods include 
\begin{itemize}
\item the Detrended Moving Average technique \cite{BMA1,BMA2,BMA3}, which we denote 
by Backward Moving Average (BMA) technique (following \cite{CMA}), 
\item the Centred Moving Average Average (CMA) method \cite{CMA}, an essentially 
improved version of BMA, 
\item the Modified Detrended Fluctuation Analysis (MDFA) \cite{MDFA}, which is 
essentially a mixture of old FA and DFA, 
\item the continuous DFA (CDFA) technique \cite{CDFA1,CDFA2}, which is particularly 
useful for the detection of crossovers, 
\item the Fourier DFA \cite{FourierDFA},
\item a variant of DFA based on empirical mode decomposition (EMD) \cite{EMD-DFA}, 
\item a variant of DFA based on singular value decomposition (SVD) 
\cite{SVD-DFA1,SVD-DFA2}, and
\item a variant of DFA based on high-pass filtering \cite{HP-DFA}.
\end{itemize}

Detrended Moving Average techniques will be thoroughly described and discussed in 
the next section.  A study comparing DFA with CMA and MDFA can be found in \cite{amir07}.  
For studies comparing DFA and BMA, see \cite{grech05,xu05}; note that \cite{xu05} also 
discusses CMA.  

The method we denote as {\it Modified Detrended Fluctuation Analysis (MDFA)} \cite{MDFA}, 
eliminates trends similar to the DFA method. A polynomial is fitted to the profile 
function $Y(j)$ in each segment $\nu$ and the deviation between the profile function and 
the polynomial fit is calculated, $\tilde{Y}_s(j) = Y(j)-y_{\nu,s}^{p}(j)$ 
(Eq.~(\ref{detrendedfunction})).  To estimate correlations in the data, this method 
uses a derivative of $\tilde{Y}_s(j)$, obtained for each segment $\nu$, by 
$\Delta\tilde{Y}_s(j)=\tilde{Y}_s(j+s/2)-\tilde{Y}_s(j)$. Hence, the fluctuation 
function (compare with Eqs.~(\ref{Fs2}) and (\ref{dfap})) is calculated as follows:
\begin{equation} F_2(s)=\left[\frac{1}{N}\sum_{j=1}^{N} \left(\tilde{Y}_s(j+s/2)-
\tilde{Y}_s(j)\right)^2 \right]^{1/2}. \label{fluctuationfunctionJFA} \end{equation}
As in case of DFA, MDFA can easily be generalised to remove higher order trends in the data.
Since the fitting polynomials in adjacent segments are not related, $\tilde{Y}_s(j)$
shows abrupt jumps on their boundaries as well. This leads to fluctuations of $F_2(s)$ for 
large segment sizes $s$ and limits the maximum usable scale to $s<N/4$ as for DFA.
The detection of crossovers in the data, however, is more exact with MDFA (compared
with DFA), since no correction of the estimated crossover time scales seems to be needed 
\cite{amir07}.

The {\it Fourier-detrended fluctuation analysis} \cite{FourierDFA} aims to eliminate 
slow oscillatory trends which are found especially in weather and climate series due to 
seasonal influences. The character of these trends can be rather periodic and regular or 
irregular, and their influence on the detection of long-range correlations by means of 
DFA was systematically studied previously \cite{kantelhardt01}. Among other things it has 
been shown that slowly varying periodic trends disturb the scaling behaviour of the results 
much stronger than quickly oscillating trends and thus have to be removed prior to the 
analysis. In case of periodic and regular oscillations, e.~g., in temperature fluctuations 
one simply removes the low frequency seasonal trend by subtracting the daily mean 
temperatures from the data. Another way, which the Fourier-detrended fluctuation 
analysis suggests, is to filter out the relevant frequencies in the signals' Fourier 
spectrum before applying DFA to the filtered signal. Nevertheless, this method faces 
several difficulties especially its limitation to periodic and regular trends and the 
need for a priori knowledge of the interfering frequency band. 

To study correlations in data with quasi-periodic or irregular oscillating trends, {\it 
empirical mode decomposition} (EMD) was suggested \cite{EMD-DFA}. 
The EMD algorithm breaks down the signal into its intrinsic mode functions (IMFs) which 
can be used to distinguish between fluctuations and background. The background, estimated 
by a quasi-periodic fit containing the dominating frequencies of a sufficiently large 
number of IMFs, is subtracted from the data, yielding a slightly better scaling behaviour 
in the DFA curves.  However, we believe that the method might be too complicated for 
wide-spread applications.

Another method which was shown to minimise the effect of periodic and quasi-periodic 
trends is based on {\it singular value decomposition} (SVD) \cite{SVD-DFA1, SVD-DFA2}. In 
this approach, one first embeds the original signal in a matrix whose dimension has to be 
much larger than the number of frequency components of the periodic or quasi-periodic 
trends obtained in the power spectrum. Applying SVD yields a diagonal matrix which can be 
manipulated by setting the dominant eigen-values (associated with the trends) to zero. 
The filtered matrix finally leads to the filtered data, and it has been shown that subsequent 
application of DFA determines the expected scaling behaviour if the embedding dimension
is sufficiently large. None the less, the performance of this rather complex method seems 
to decrease for larger values of the scaling exponent. Furthermore SVD-DFA assumes that 
trends are deterministic and narrow banded.

The detrending procedure in DFA (Eq.~(\ref{detrendedfunction})) can be regarded as a 
scale-dependent high-pass filter since (low-frequency) fluctuations exceeding a specific 
scale $s$ are eliminated. Therefore, it has been suggested to obtain the detrended profile 
$\tilde{Y}_s(j)$ for each scale $s$ directly by applying digital high-pass filters 
\cite{HP-DFA}. In particular, Butterworth, Chebyshev-I, Chebyshev-II, and an elliptical 
filter were suggested. While the elliptical filter showed the best performance in detecting 
long-range correlations in artificial data, the Chebyshev-II filter was found to be 
problematic. Additionally, in order to avoid a time shift between filtered and original 
profile, the average of the directly filtered signal and the time reversed filtered signal 
is considered. The effects of these complicated filters on the scaling behaviour are, 
however, not fully understood. 

Finally, a continuous DFA method has been suggested in the context of studying heartbeat
data during sleep \cite{CDFA1,CDFA2}.  The method compares unnormalised fluctuation functions 
$F_2(s)$ for increasing length of the data.  I.~e., one starts with a very short recording and
subsequently adds more points of data.  The method is particularly suitable for the detection
of change points in the data, e.~g., physiological transitions between different activity or
sleep stages.  Since the main objective of the method is not the study of scaling behaviour,
we do not discuss it in detail here.

\subsection{Centered Moving Average (CMA) Analysis}

Particular attractive modifications of DFA are the {\it Detrended Moving Average} 
(DMA) methods, where running averages replace the polynomial fits. The first suggested 
version, the {\it Backward Moving Average} (BMA) method \cite{BMA1,BMA2,BMA3}, however, 
suffers from severe problems, because an artificial time shift of $s$ between the 
original signal and the moving average is introduced.  This time shift leads to an 
additional contribution to the detrended profile $\tilde Y_s(j)$, which causes a larger 
fluctuation function $F_2(s)$ in particular for small scales in the case of long-term 
correlated data.  Hence, the scaling exponent $\alpha$ is systematically underestimated 
\cite{grech05}.  In addition, the BMA method preforms even worse for data with trends 
\cite{xu05}, and its slope is limited by $\alpha<1$ just as for the non-detrending 
method FA.

It was soon recognised that the intrinsic error of BMA can be overcome by eliminating 
the artificial time shift. This leads to the {\it Centred Moving Average} (CMA) method 
\cite{CMA}, where $\tilde{Y}_s(j)$ is calculated as 
\begin{equation} \tilde{Y}_s(j)=Y(j)-\frac{1}{s}\sum_{i=-(s-1)/2}^{(s-1)/2}Y(j+i),
\label{CMA} \end{equation}
replacing Eq.~(\ref{detrendedfunction}) while Eq.~(\ref{dfap}) and the rest of the
DFA procedure described in Section 5.3 stay the same.  Unlike DFA, the CMA method 
cannot easily be generalised to remove linear and higher order trends in the data. 

It was recently proposed \cite{amir07} that the scaling behaviour of the CMA method is 
more stable than for DFA1 and MDFA1, suggesting that CMA could be used for reliable 
computation of $\alpha$ even for scales $s<10$ (without correction of any systematic
deviations needed in DFA for this regime) and up to $s_{\rm max} = N/2$.  The standard 
errors in determining the scaling exponent $\alpha$ by fitting straight lines to the 
double logarithmic plots of $F_2(s)$ have been studied in \cite{amir07}; they are 
comparable with DFA1 (see end of Section 5.3).

Regarding the determination of crossovers, CMA is comparable to DFA1.  Ultimately, the 
CMA seems to be a good 
alternative to DFA1 when analysing the scaling properties in short data sets without 
trends. Nevertheless for data with possible unknown trends we recommend the application 
of standard DFA with several different detrending polynomial orders in order to 
distinguish real crossovers from artificial crossovers due to trends. In addition, an 
independent approach (e.~g., wavelet analysis) should be used to confirm findings of 
long-term correlations (see also Section 5.4).

\section{Methods for Multifractal Time Series Analysis}

This chapter describes the multifractal characterisation of time series, for an
introduction, see Section 3.4.  The simplest type of multifractal analysis is based 
upon the standard partition function multifractal formalism, which has been developed 
for the multifractal characterisation of normalised, stationary measures 
\cite{feder88,peitgen04,barabasi,bacry01}.  Unfortunately, this standard formalism 
does not give correct results for non-stationary time series that are affected by trends 
or that cannot be normalised. Thus, in the early 1990s an improved multifractal 
formalism has been developed, the wavelet transform modulus maxima (WTMM) method 
\cite{wtmm,wtmm1,wtmm2,wtmm2a,wtmm3}, which is based on wavelet analysis and involves 
tracing the maxima lines in the continuous wavelet transform over all scales.  An 
important alternative is the multifractal DFA (MF-DFA) algorithm \cite{kantelhardt02}, 
which does not require the modulus maxima procedure, and hence involves little more 
effort in programming than the conventional DFA. For studies comparing methods for 
detrending multifractal analysis (multifractal DFA (MF-DFA) and wavelet transform 
modulus maxima (WTMM) method), see \cite{kantelhardt02,kantelhardt03,oswiecimka06}.

\subsection{The Structure Function Approach and Singularity Spectra}

In the general multifractal formalism, one considers a normalised measure $\mu(t)$,
$t \in [0,1]$, and defines the box probabilities $\tilde \mu_s(t)=\int_{t-s/2}^{t+s/2} 
\mu(t')\, dt'$ in neighbourhoods of (scale) length $s \ll 1$ around $t$.  The 
multifractal approach is then introduced by the partition function
\begin{equation} Z_q(s)=\sum_{\nu=0}^{1/s-1} \tilde \mu_s^q[(\nu+1/2)s] \sim 
s^{\tau(q)} \quad {\rm for} \quad s \ll 1, \label{Zqkont} \end{equation}
where $\tau(q)$ is the Renyi scaling exponent and $q$ is a real parameter that can
take positive as well as negative moments.  Note that $\tau(q)$ is sometimes defined 
with opposite sign (see, e.~g., \cite{feder88}).  A record is called monofractal (or 
self-affine), when the Renyi scaling exponent $\tau(q)$ depends linearly on 
$q$; otherwise it is called multifractal.  The generalised multifractal dimensions
$D(q)$ (see also Section 3.4) are related to $\tau(q)$ by $D(q)=\tau(q)/{q-1}$,
such that the fractal dimension of the support is $D(0)=-\tau(0)$ and the correlation
dimension is $D(2)=\tau(2)$.

In time series, a discrete version has to be used, and the considered data $(x_i)$, 
$i=1, \ldots, N$ may usually include negative values.  Hence, setting $N_s={\rm 
int}(N/s)$ and $X(\nu,s)=\sum_{i=1}^s x_{\nu s+i}$ for $\nu=0,\ldots,N_s-1$ we can 
define \cite{feder88,peitgen04}, 
\begin{equation} Z_q(s) = \sum_{\nu=0}^{N_s-1} \vert X(\nu,s)\vert^q \sim s^{\tau(q)} 
\quad {\rm for} \quad s>1. \label{Zq} \end{equation}
Inserting the profile $Y(j)$ and $F_{\rm FA}(\nu,s)$ from Eqs.~(\ref{profile}) and 
(\ref{fa2}), respectively, we obtain
\begin{equation} Z_q(s) = \sum_{\nu=0}^{N_s-1} \left\{[Y((\nu+1) s) - Y(\nu s)]^2
\right\}^{q/2} = \sum_{\nu=0}^{N_s-1} F_{\rm FA}^{q/2}(\nu,s). \label{ZqF} \end{equation}
Comparing Eqs.~(\ref{ZqF}) with (\ref{Fs2}), we see that this multifractal approach 
can be considered as a generalised version of the Fluctuation Analysis (FA) method, 
where the exponent 2 is replaced by $q$.  In particular we find (disregarding the 
summation over the second partition of the time series)
\begin{equation} F_2(s) \sim \left[ {1 \over N_s} Z_2(s) \right]^{1/2} \sim 
s^{[1+\tau(2)]/2} \quad \Rightarrow \quad 2\alpha = 1+\tau(2)=1+D(2).
\label{Z2F2} \end{equation}
We thus see that all methods for (mono-)fractal time analysis (discussed in Chapters 
4 and 5) in fact study the correlation dimension $D(2) = 2\alpha-1 = \beta = 1-\gamma$
(see Eq.~(\ref{exponents})).  

It is straightforward to define a generalised (multifractal) Hurst exponent $h(q)$ for 
the scaling behaviour of the $q$th moments of the fluctuations \cite{barabasi,bacry01},
\begin{equation} F_q(s) = \left[{1\over N_s}Z_2(s)\right]^{1/q} \sim s^{[1+\tau(q)]/q} 
=s^{h(q)} \quad \Rightarrow \quad h(q)= {1+\tau(q) \over q}. \label{tauH} \end{equation}
with $h(2) = \alpha \approx H$.  In the following, we will use only $h(2)$ for the 
standard fluctuation exponent (denoted by $\alpha$ in the previous chapters), and 
reserve the letter $\alpha$ for the H\"older exponent.

Another way to characterise a multifractal series is the singularity spectrum $f(\alpha)$, 
that is related to $\tau(q)$ via a Legendre transform \cite{feder88,peitgen04},
\begin{equation} \alpha = {d \over dq} \tau(q) \quad {\rm and} \quad
f(\alpha) = q \alpha - \tau(q). \label{Legendre} \end{equation}
Here, $\alpha$ is the singularity strength or H\"older exponent (see also articles
... in the encyclopedia), while $f(\alpha)$ denotes the dimension of the subset of 
the series that is characterised by $\alpha$.  Note that $\alpha$ is {\it not} the 
fluctuation scaling exponent in this section, although the same letter is traditionally 
used for both.  Using Eq.~(\ref{tauH}), we can directly relate $\alpha$ and $f(\alpha)$ 
to $h(q)$,
\begin{equation} \alpha = h(q) + q h'(q) \quad {\rm and} \quad
f(\alpha) = q [\alpha - h(q)] + 1.\label{Legendre2} \end{equation}

\subsection{Wavelet Transform Modulus Maxima (WTMM) Method}

The wavelet transform modulus maxima (WTMM) method \cite{wtmm,wtmm1,wtmm2,wtmm2a,wtmm3} 
is a well-known method to investigate the multifractal scaling properties of fractal and 
self-affine objects in the presence of non-stationarities. For applications, see e.~g.
\cite{wtmmanw,wtmmanw1}.  It is based upon the wavelet transform with continuous basis 
functions as defined in Section 5.1, Eq.~(\ref{wt}).  Note that in this case the series 
$\tilde x_i$ are analysed directly instead of the profile $Y(j)$ defined in 
Eq.~(\ref{profile}).  Using wavelets orthogonal to $m$th order polynomials, the 
corresponding trends are elliminated.

Instead of averaging over all wavelet coefficients $L_\psi(\tau,s)$, one averages,
within the modulo-maxima method, only the local maxima of $\vert L_\psi(\tau,s) 
\vert$.  First, one determines for a given scale $s$, the positions $\tau_j$ of the 
local maxima of $\vert W(\tau,s) \vert$ as function of $\tau$, so that $\vert 
L_\psi(\tau_j-1,s) \vert < \vert L_\psi(\tau_j,s) \vert \ge \vert L_\psi(\tau_j+1,s) 
\vert$ for $j=1,\ldots,j_{\rm max}$. This maxima procedure is demonstrated in 
Fig.~{\fig 7}.  Then one sums up the $q$th power of the maxima,
\begin{equation} Z(q,s) = \sum_{j=1}^{j_{\rm max}} \vert L_\psi(\tau_j,s)
\vert^q. \label{eq11a}\end{equation}
The reason for the maxima procedure is that the absolute wavelet coefficients $\vert 
L_\psi(\tau,s) \vert$ can become arbitrarily small.  The analysing wavelet $\psi(x)$ 
must always have positive values for some $x$ and negative values for other $x$, since 
it has to be orthogonal to possible constant trends.  Hence there are always positive 
and negative terms in the sum (\ref{wt}), and these terms might cancel. If that 
happens, $\vert L_\psi(\tau,s) \vert$ can become close to zero. Since such small 
terms would spoil the calculation of negative moments in Eq.~(\ref{eq11a}), they have 
to be eliminated by the maxima procedure.  

In fluctuation analysis, on the other hand, the calculation of the variances $F^2(\nu,s)$, 
e.~g. in Eq.~(\ref{fa2}), involves only positive terms under the summation.  The variances 
cannot become arbitrarily small, and hence no maximum procedure is required for series 
with compact support.  In addition, the variances will always increase if the segment 
length $s$ is increased, because the fit will always be worse for a longer segment.  In 
the WTMM method, in contrast, the absolute wavelet coefficients $\vert L_\psi(\tau,s) 
\vert$ need not increase with increasing scale $s$, even if only the local maxima are 
considered. The values $\vert L_\psi(\tau,s) \vert$ might become smaller for increasing 
$s$ since just more (positive and negative) terms are included in the summation 
(\ref{wt}), and these might cancel even better.  Thus, an additional supremum procedure 
has been introduced in the WTMM method in order to keep the dependence of $Z(q,s)$ on 
$s$ monotonous.  If, for a given scale $s$, a maximum at a certain position $\tau_j$ 
happens to be smaller than a maximum at $\tau'_j \approx \tau_j$ for a lower scale $s' 
< s$, then $L_\psi(\tau_j,s)$ is replaced by $L_\psi(\tau'_j,s')$ in Eq.~(\ref{eq11a}).  

Often, scaling behaviour is observed for $Z(q,s)$, and scaling exponents $\hat{\tau}(q)$ 
can be defined that describe how $Z(q,s)$ scales with $s$,
\begin{equation} Z(q,s) \sim s^{\hat{\tau}(q)}. \label{eq11b}\end{equation}
The exponents $\hat{\tau}(q)$ characterise the multifractal properties of the series 
under investigation, and theoretically they are identical with the $\tau(q)$ defined 
in Eq.~(\ref{Zq}) \cite{wtmm,wtmm1,wtmm2,wtmm3} and related to $h(q)$ by Eq.~(\ref{tauH}).

\subsection{Multifractal Detrended Fluctuation Analysis (MF-DFA)}

The multifractal DFA (MF-DFA) procedure consists of five steps \cite{kantelhardt02}.  
The first three steps are essentially identical to the conventional DFA procedure (see 
Section 5.3 and Fig.~{\fig 4}).  Let us assume that $(\tilde x_i)$ is a series of 
length $N$, and that this series is of compact support.  The support can be defined as 
the set of the indices $j$ with nonzero values $\tilde x_j$, and it is compact if $\tilde 
x_j = 0$ for an insignificant fraction of the series only.  The value of $\tilde x_j=0$ 
is interpreted as having no value at this $j$. Note that we are {\it not} discussing the 
fractal or multifractal features of the plot of the time series in a two-dimensional 
graph (see also the discussion in Section 3.1), but analysing time series as 
one-dimensional structures with values assigned to each point.  Since real time 
series always have finite length $N$, we explicitly want to determine the 
multifractality of finite series, and we are not discussing the limit for $N \to 
\infty$ here (see also Section 6.1). 

\begin{itemize}
\item {\it Step 1}: Calculate the profile $Y(j$, Eq.~(\ref{profile}), by 
integrating the time series. \\
\item {\it Step 2}: Divide the profile $Y(j)$ into $N_s = {\rm int}(N/s)$ 
non-overlapping segments of equal length $s$. Since the length $N$ of the series is 
often not a multiple of the considered time scale $s$, the same procedure can be 
repeated starting from the opposite end.  Thereby, $2 N_s$ segments are obtained 
altogether. \\
\item {\it Step 3}: Calculate the local trend for each of the $2 N_s$ 
segments by a least-square fit of the profile.  Then determine the variance by Eqs.
(\ref{detrendedfunction}) and (\ref{dfap}) for each segment $\nu=0,\ldots,2N_s-1$
Again, linear, quadratic, cubic, or higher order polynomials can be used in the 
fitting procedure, and the corresponding methods are thus called MF-DFA1, MF-DFA2, 
MF-DFA3, $\ldots$) \cite{kantelhardt02}. In (MF-)DFA$m$ [$m$th order (MF-)DFA] 
trends of order $mp$ in the profile (or, equivalently, of order $m-1$ in the original
series) are eliminated.  Thus a comparison of the results for different orders of 
DFA allows one to estimate the type of the polynomial trend in the time series  
\cite{kantelhardt01,hu01}.\\
\item {\it Step 4}: Average over all segments to obtain the $q$th order fluctuation 
function
\begin{equation} F_q(s) = \left\{ {1 \over 2 N_s} \sum_{\nu=1}^{2 N_s} \left[ 
F_{{\rm DFA}m}^2(\nu,s) \right]^{q/2} \right\}^{1/q}, \label{fdef}\end{equation}
This is the generalization of Eq.~(\ref{Fs2}) suggested by the relations derived in 
Section 6.1.  For $q=2$, the standard DFA procedure is retrieved.  One is interested 
in how the generalised $q$ dependent fluctuation functions $F_q(s)$ depend on the 
time scale $s$ for different values of $q$.  Hence, we must repeat steps 2 to 4 for
several time scales $s$.  It is apparent that $F_q(s)$ will increase with increasing 
$s$.  Of course, $F_q(s)$ depends on the order $m$.  By construction, $F_q(s)$ 
is only defined for $s \ge m+2$. \\
\item {\it Step 5}: Determine the scaling behaviour of the fluctuation functions 
by analysing log-log plots $F_q(s)$ versus $s$ for each value of $q$. If the series 
$\tilde x_i$ are long-range power-law correlated, $F_q(s)$ increases, for large values 
of $s$, as a power-law,
\begin{equation} F_q(s) \sim s^{h(q)} \quad  {\rm with} \quad h(q)= {1+\tau(q) 
\over q}. \label{Hq} \end{equation}
\end{itemize}

For very large scales, $s>N/4$, $F_q(s)$ becomes statistically unreliable because 
the number of segments $N_s$ for the averaging procedure in step 4 becomes very small.  
Thus, scales $s>N/4$ should be excluded from the fitting procedure determining $h(q)$.
Besides that, systematic deviations from the scaling behaviour in Eq.~(\ref{Hq}), which 
can be corrected, occur for small scales $s \approx 10$. 

The value of $h(0)$, which corresponds to the limit $h(q)$ for $q \to 0$, cannot be 
determined directly using the averaging procedure in Eq.~(\ref{fdef}) because of the 
diverging exponent.  Instead, a logarithmic averaging procedure has to be employed,
\begin{equation} F_0(s) = \exp \left\{ {1 \over 4 N_s} \sum_{\nu=1}^{2 N_s} 
\ln \left[F^2(\nu,s)\right] \right\} \sim s^{h(0)}. \end{equation}
Note that $h(0)$ cannot be defined for time series with fractal support, where $h(q)$ 
diverges for $q \to 0$.

For monofractal time series with compact support, $h(q)$ is independent of $q$, since 
the scaling behaviour of the variances $F^2_{{\rm DFA}m}(\nu,s)$ is identical for all 
segments $\nu$, and the averaging procedure in Eq.~(\ref{fdef}) will give just this 
identical scaling behaviour for all values of $q$.  Only if small and large fluctuations
scale differently, there will be a significant dependence of $h(q)$ on $q$:  If we 
consider positive values of $q$, the segments $\nu$ with large variance $F^2(\nu,s)$ 
(i.~e. large deviations from the corresponding fit) will dominate the average $F_q(s)$.  
Thus, for positive values of $q$, $h(q)$ describes the scaling behaviour of the segments 
with large fluctuations.  On the contrary, for negative values of $q$, the segments 
$\nu$ with small variance $F^2_{{\rm DFA}m}(\nu,s)$ will dominate the average $F_q(s)$.  
Hence, for negative values of $q$, $h(q)$ describes the scaling behaviour of the segments 
with small fluctuations.  Figure {\fig 8} shows typical results obtained for $F_q(s)$
in the MF-DFA procedure.

Usually the large fluctuations are characterised by a smaller scaling exponent $h(q)$ 
for multifractal series than the small fluctuations. This can be understood from the 
following arguments:  For the maximum scale $s=N$ the fluctuation function $F_q(s)$ 
is independent of $q$, since the sum in Eq.~(\ref{fdef}) runs over only two identical 
segments.  For smaller scales $s \ll N$ the averaging procedure runs over several 
segments, and the average value $F_q(s)$ will be dominated by the $F^2(\nu,s)$ from 
the segments with small (large) fluctuations if $q<0$ ($q>0$).  Thus, for $s \ll N$,
$F_q(s)$ with $q < 0$ will be smaller than $F_q(s)$ with $q > 0$, while both become 
equal for $s=N$.  Hence, if we assume an homogeneous scaling behaviour of $F_q(s)$ 
following Eq.~(\ref{Hq}), the slope $h(q)$ in a log-log plot of $F_q(s)$ with $q<0$ 
versus $s$ must be larger than the corresponding slope for $F_q(s)$ with $q > 0$.  
Thus, $h(q)$ for $q<0$ will usually be larger than $h(q)$ for $q>0$.

However, the MF-DFA method can only determine {\it positive} generalised Hurst 
exponents $h(q)$, and it already becomes inaccurate for strongly anti-correlated 
signals when $h(q)$ is close to zero. In such cases, a modified (MF-)DFA technique 
has to be used.  The most simple way to analyse such data is to integrate the time 
series before the MF-DFA procedure.  Following the MF-DFA procedure as described 
above, we obtain a generalised fluctuation functions described by a scaling law 
with $\tilde h(q) = h(q)+1$.  The scaling behaviour can thus be accurately determined 
even for $h(q)$ which are smaller than zero for some values of $q$.

The accuracy of $h(q)$ determined by MF-DFA certainly depends on the length $N$ of 
the data.  For $q=\pm 10$ and data with $N=$ 10 000 and 100 000, systematic and 
statistical error bars (standard deviations) up to $\Delta h(q) \approx \pm 0.1$ and
$\approx \pm 0.05$ should be expected, respectively \cite{kantelhardt02}.  A 
difference of $h(-10) - h(+10) = 0.2$, corresponding to an even larger width 
$\Delta \alpha$ of the singularity spectrum $f(\alpha)$ defined in Eq.~(\ref{Legendre}) 
is thus not significant unless the data was longer than $N=$ 10 000 points.  Hence, 
one has to be very careful when concluding multifractal properties from differences 
in $h(q)$.

As already mentioned in the introduction, two types of multifractality in time series 
can be distinguished.  Both of them require a multitude of scaling exponents for small 
and large fluctuations:  (i) Multifractality of a time series can be due to a broad 
probability density function for the values of the time series, and (ii) 
multifractality can also be due to different long-range correlations for small and 
large fluctuations.  The most easy way to distinguish between these two types is by 
analysing also the corresponding randomly shuffled series \cite{kantelhardt02}.  In 
the shuffling procedure the values are put into random order, and thus all correlations 
are destroyed.  Hence the shuffled series from multifractals of type (ii) will exhibit 
simple random behaviour, $h_{\rm shuf}(q) = 0.5$, i.~e. non-multifractal scaling.  
For multifractals of type (i), on the contrary, the original $h(q)$ dependence is not 
changed, $h(q)=h_{\rm shuf}(q)$, since the multifractality is due to the probability 
density, which is not affected by the shuffling procedure.  If both kinds of 
multifractality are present in a given series, the shuffled series will show weaker 
multifractality than the original one.

\subsection{Comparison of WTMM and MF-DFA}

The MF-DFA results turn out to be slightly more reliable than the WTMM results
\cite{kantelhardt02,kantelhardt03,oswiecimka06}.  In particular, the MF-DFA has
slight advantages for negative $q$ values and short series.  In the other cases 
the results of the two methods are rather equivalent.  Besides that, the main 
advantage of the MF-DFA method compared with the WTMM method lies in the simplicity 
of the MF-DFA method.  However, contrary to WTMM, MF-DFA is restricted to studies 
of data with full one-dimensional support, while WTMM is not.  Both, WTMM and 
MF-DFA have been generalised for higher dimensional data, see \cite{gu06} for 
higher dimensional MF-DFA and, e.~g., \cite{wtmm3} for higher dimensional WTMM.
Studies of other generalisations of detrending methods like the discrete WT approach 
(see Section 5.2) and the CMA method (see Section 5.7) are currently under 
investigation \cite{mfcma}.

\section{Statistics of Extreme Events in Fractal Time Series}

The statistics of return intervals between well defined extremal events is a powerful
tool to characterise the temporal scaling properties of observed time series and
to derive quantities for the estimation of the risk for hazardous events like floods,
very high temperatures, or earthquakes.  It was shown recently that long-term 
correlations represent a natural mechanism for the clustering of the hazardous 
events \cite{PRL}.  In this chapter we will discuss the most important consequences
of long-term correlations and fractal scaling of time series upon the statistics
of extreme events \cite{PRL,physA_I,kantz,PRE1,PRE2}.  Corresponding work 
regarding multifractal data \cite{bogachev07} is not discussed here.

\subsection{Return Intervals Between Extreme Events}

To study the statistics of return intervals we consider again a time series $(x_i)$, 
$i=1,\ldots,N$ with fractal scaling behaviour, sampled homogeneously and normalised 
to zero mean and unit variance.  For describing the reoccurrence of rare events 
exceeding a certain threshold $q$, we investigate the return intervals $r_q$ between 
these events, see Fig.~{\fig 9}. The average return interval $R_q = \langle r_q 
\rangle$ increases as a function of the threshold $q$ (see, e.~g. 
\cite{vStorch+Zwiewers}).  It is known that for uncorrelated records ('white noise'), 
the return intervals are also uncorrelated and distributed according to the Poisson 
distribution, $P_q(r) = \frac{1}{R_q} \exp(-r/R_q)$.  For fractal (long-term correlated) 
data with auto-correlations following Eq.~(\ref{lrc}), we obtain a {\it stretched} 
exponential \cite{PRL,physA_I,kantz,PRE1,Rosenblatt},
\begin{equation} P_q(r) = {a_\gamma \over R_q} \exp [ -b_\gamma (r/R_q)^\gamma], 
\quad \label{stretched} \end{equation}

This behaviour is shown in Fig.~{\fig 10}.
The exponent $\gamma$ is the correlation exponent from $C(s)$, and the parameters
$a_\gamma$ and $b_\gamma$ are independent of $q$.  They can be determined from 
the normalization conditions for $P_q(r)$, i.~e., $\int P_q(r) \, dr = 1$ and
$\int rP_q(r) \, dr = R_q$.  The form of the distribution (\ref{stretched}) 
indicates that return intervals both well below and well above their average 
value $R_q$ (which is independent of $\gamma$) are considerably more frequent 
for long-term correlated than for uncorrelated data.  It has to be noted that there
are deviations from the stretched exponential law (\ref{stretched}) for very small
$r$ (discretization effects and an additional power-law regime) and for very large 
$r$ (finite size effects), see Fig.~{\fig 10}.  The extent of the deviations from 
Eq.~(\ref{stretched}) depends on the distribution of the values $x_i$ of the time 
series. For a discussion of these effects, see \cite{PRE1}. 

Equation (\ref{stretched}) does not quantify, however, if the return intervals themselves 
are arranged in a correlated or in an uncorrelated fashion, and if clustering of rare 
events may be induced by long-term correlations.  To study this question, one has to
evaluate the auto-correlation function $C_r(s) = \langle r_q(l)r_q(l+s) \rangle-R_q^2$ 
of the return intervals. The results for model data suggests that also the return 
intervals are long-term power-law correlated, with the same exponent $\gamma$ as the 
original record.  Accordingly, large and small return intervals are not arranged in a 
random fashion but are expected to form clusters.  As a consequence, the probability of 
finding a certain return interval $r$ depends on the value of the preceding interval 
$r_0$, and this effect has to be taken into account in predictions and risk estimations
\cite{PRL,PRE1}. 

The conditional distribution function $P_q (r\vert r_0)$ is a basic quantity, from which 
the relevant quantities in risk estimations can be derived \cite{vStorch+Zwiewers}.  For 
example, the first moment of $P_q (r\vert r_0)$ is the average value $R_q(r_0)$ of those 
return intervals that directly follow $r_0$. By definition, $R_q(r_0)$ is the expected 
waiting time to the next event, when the two events before were separated by $r_0$. The 
more general quantity is the expected waiting time $\tau_q(x\vert r_0)$ to the next 
event, when the time $x$ has been elapsed. For $x=0$, $\tau_q(0\vert r_0)$ is identical 
to $R_q(r_0)$. In general, $\tau_q(x\vert r_0)$ is related to $P_q(r\vert r_0)$ by

\begin{equation} \tau_q(x\vert r_0)=\int_x^\infty (r-x) P_q(r\vert r_0)dr/
\int_x^\infty P_q(r\vert r_0)dr \end{equation}

For uncorrelated records, $\tau_q(x\vert r_0)/R_q = 1$ (except for discreteness 
effects that lead to $\tau_q(x\vert r_0)/R_q > 1$ for $x > 0$, see \cite{Sornette}). 
Due to the scaling of $P_q(r\vert r_0)$, also $\tau_q(x\vert r_0)/R_q$ scales with 
$r_0/R_q$ and $x/R_q$.  Small and large return intervals are more likely to be 
followed by small and large ones, respectively, and hence $\tau_q(0\vert r_0)/R_q =
R_q(r_0)/R_q$ is well below (above) one for $r_0/R_q$ well below (above) one. With 
increasing $x$, the expected residual time to the next event increases. Note that only 
for an infinite long-term correlated record, the value of $\tau_q(x\vert r_0)$ will 
increase indefinitely with $x$ and $r_0$. For real (finite) records, there exists 
a maximum return interval which limits the values of $x$, $r_0$ and $\tau_q(x\vert 
r_0)$.

\subsection{Distribution of Extreme Events}

In this section we describe how the presence of fractal long-term correlations affects 
the statistics of the extreme events, i.~e., maxima within time segments of fixed 
duration $R$, see Fig.~{\fig 11} for illustration.  By definition, extreme events are 
rare occurrences of extraordinary nature, such as, e.~g. floods, very high temperatures, 
or earthquakes.  In hydrological engineering such conventional extreme value statistics 
are commonly applied to decide what building projects are required to protect riverside 
areas against typical floods that occur for example once in 100 years.  Most of these 
results are based on statistically independent values $x_i$ and hold only in the limit 
$R \to \infty$.  However, both of these assumptions are not strictly fulfilled for 
correlated fractal scaling data.

In classical extreme value statistics one assumes that records $(x_i)$ consist of i.i.d. 
data, described by density distributions $P(x)$, which can be, e.~g., a Gaussian or an 
exponential distribution.  One is interested in the distribution density function 
$P_R(m)$ of the maxima $(m_j)$ determined in segments of length $R$ in the original 
series $(x_i)$, see Fig.~{\fig 11}.  Note that all maxima are also elements of the 
original data.  The corresponding integrated maxima distribution $G_R(m)$ is defined as
\begin{equation}
G_R(m) = 1- E_R(m) = \int_{-\infty}^{m} P_R(m') \; dm'.
\label{eq:eichner:integrate}\end{equation}
Since $G_R(m)$ is the probability of finding a maximum smaller than $m$, $E_R(m)$
denotes the probability of finding a maximum that exceeds $m$.  One of the main
results of traditional extreme value statistics states that for independently and
identically distributed (i.i.d.) data $(x_i)$ with Gaussian or exponential
distribution density function $P(x)$ the integrated distribution $G_R(m)$ converges
to a double exponential (Fisher-Tippet-Gumbel) distribution (often labelled as Type I) 
\cite{fisher,gumbel,galambos2,leadbetter,galambos}, i.~e.,
\begin{equation}
G_R(m) \to G\left({m - u \over \alpha} \right) = \exp
\left[ -\exp \left(-{m - u \over \alpha}\right) \right]
\label{eq:eichner:gumbel}\end{equation}
for $R \to \infty$, where $\alpha$ is the scale parameter and $u$ the location parameter.
By the method of moments those parameters are given by $\alpha = \frac{\sqrt{6}}{\pi}
\sigma_R$ and $u = m_R - n_e \alpha$ with the Euler constant $n_e = 0.577216$
\cite{leadbetter,chow,raud,rasmussen}.  Here $m_R$ and $\sigma_R$ denote the 
($R$-dependent) mean maximum and the standard deviation, respectively.  Note that 
different asymptotics will be reached for broader distributions of data $(x_i)$ that 
belong to other domains of attraction \cite{leadbetter}. For example, for data following 
a power-law distribution (or Pareto distribution), $P(x) = (x/x_0)^{-k}$, $G_R(m)$ 
converges to a Fr\'echet distribution, often labelled as Type II.  For data following a 
distribution with finite upper endpoint, for example the uniform distribution $P(x) = 1$ 
for $0 \le x \le 1$ , $G_R(m)$ converges to a Weibull distribution, often labelled as 
Type III.  We do not consider the latter two types of asymptotics here.

Numerical studies of fractal model data have recently shown that the distribution $P(x)$ 
of the original data has a much stronger effect upon the convergence towards the Gumbel 
distribution than the long-term correlations in the data.  Long-term correlations just 
slightly delay the convergence of $G_R(m)$ towards the Gumbel distribution (\ref
{eq:eichner:gumbel}).  This can be observed very clearly in a plot of the integrated and 
scaled distribution $G_R(m)$ on logarithmic scale \cite{PRE2}.

Furthermore, it was found numerically that (i) the maxima series $(m_j)$ exhibit long-term 
correlations similar to those of the original data $(x_i)$, and most notably (ii) the 
maxima distribution as well as the mean maxima significantly depend on the history, in 
particular on the previous maximum \cite{PRE2}.  The last item implies that conditional 
mean maxima and conditional maxima distributions should be considered for improved extreme 
event predictions.

\section{Simple Models for Fractal and Multifractal Time Series}

\subsection{Fourier Filtering}

Fractal scaling with long-term correlations can be introduced most easily into time 
series by the Fourier-filtering technique, see, e.~g., \cite{Mandelbrot71,Voss85,makse}. 
The Fourier filtering technique is not limited to the generation of long-term correlated 
data characterised by a power-law auto-correlation function $C(s) \sim x^{-\gamma}$ 
with $0 < \gamma < 1$.  All values of the scaling exponents $\alpha = h(2) 
\approx H$ or $\beta = 2\alpha-1$ can be obtained, even those that cannot be found 
directly by the fractal analysis techniques described in Chapters 4 and 5 (e.~g. 
$\alpha < 1$). Note, however, that Fourier filtering will always yield Gaussian 
distributed data values and that no nonlinear or multifractal properties can be 
achieved (see also Sections 3.4, 5.5, and Chapter 6).  In Section 5.4, we have briefly 
described a modification of Fourier filtering for obtaining reliable short-term 
correlated data. 

For the generation of data characterised by fractal scaling with $\beta = 2\alpha-1$ 
\cite{Mandelbrot71,Voss85} we start with uncorrelated Gaussian distributed random numbers 
$x_i$ from an i.i.d. generator. Transforming a series of such numbers into frequency 
space with discrete Fourier transform or FFT (for suitable series lengths $N$) yields a 
flat power spectrum, since random numbers correspond to white noise.  Multiplying the 
(complex) Fourier coefficients by $f^{-\beta/2}$, where $f \propto 1/s$ is the frequency, 
will rescale the power spectrum $S(f)$ to follow Eq.~(\ref{spectr}), as expected for time 
series with fractal scaling.  After transforming back to the time domain (using inverse 
Fourier transform or inverse FFT) we will thus obtain the desired long-term correlated 
data $\tilde x_i$. The final step is the normalization of this data.  

The Fourier filtering method can be improved using modified Bessel functions instead
of the simple factors $f^{-\beta/2}$ in modifying the Fourier coefficients \cite{makse}.
This way problems with the divergence of the autocorrelation function $C(s)$ at $s=0$
can be avoided.  

An alternative method to the Fourier filtering technique, the random midpoint displacement
method, is based on the construction of self-affine surfaces by an iterative procedure, 
see, e.~g. \cite{feder88}.  Starting with one interval with constant values, the intervals 
are iterative split in the middle and the midpoint is displaced by a random offset.  The 
amplitude of this offset is scaled according to the length of the interval.  Since the 
method generates a self-affine surface $x_i$ characterised by a Hurst exponent $H$, the 
differentiated series $\Delta x_i$ can be used as long-term correlated or anti-correlated 
random numbers.  Note, however, that the correlations do not persist for the whole length 
of the data generated this way.  Another option is the use of wavelet synthesis, the 
reverse of wavelet analysis described in Section 5.1.  In that method, the scaling law is 
introduced by setting the magnitudes of the wavelet coefficients according to the 
corresponding time scale $s$.

\subsection{The Schmitz-Schreiber Method}

When long-term correlations in random numbers are introduced by the Fourier-filtering 
technique (see previous section), the original distribution $P(x)$ of the time series 
values $x_i$ is always modified such that it becomes closer to a Gaussian.  Hence, no
series $(x_i)$ with broad distributions of the values {\it and} fractal scaling 
can be generated.  In these cases an iterative algorithm introduced by Schreiber and 
Schmitz \cite{schreiber,schreiber2} must be applied.  

The algorithm consists of the following steps:  First one creates a Gaussian distributed 
long-term correlated data set with the desired correlation exponent $\gamma$ by standard 
Fourier-filtering \cite{makse}. The power spectrum $S_{\rm G}(f) = F_{\rm G}(f) 
F_{\rm G}^*(f)$ of this data set is considered as reference spectrum (where $f$ 
denotes the frequency in Fourier space and the $F_{\rm G}(f)$ are the complex Fourier 
coefficients).  Next one creates an uncorrelated sequence of random numbers $(x^{\rm 
ref}_i)$, following a desired distribution $P(x)$.  The (complex) Fourier transform 
$F(f)$ of the $(x^{\rm ref}_i)$ is now divided by its absolute value and multiplied 
by the square root of the reference spectrum, 
\begin{equation}
F_{\rm new}(f) = {F(f) \sqrt{S_{\rm G}(f)} \over \vert F(f) \vert}. 
\label{schreiber}\end{equation}
After the Fourier back-transformation of $F_{\rm new}(f)$, the new sequence 
$(x^{\rm new}_i)$ has the desired correlations (i.~e. the desired $\gamma$), but the 
shape of the distribution has changed towards a (more or less) Gaussian distribution. 
In order to enforce the desired distribution, we exchange the $(x^{\rm new}_i)$ 
by the $(x^{\rm ref}_i)$, such that the largest value of the new set is replaced by the 
largest value of the reference set, the second largest of the new set by the second 
largest of the reference set and so on.  After this the new sequence has the desired 
distribution and is clearly correlated.  However, due to the exchange algorithm the 
perfect long-term correlations of the new data sequence were slightly altered again.  
So the procedure is repeated:  the new sequence is Fourier transformed followed by 
spectrum adjustment, and the exchange algorithm is applied to the Fourier back-transformed 
data set.  These steps are repeated several times, until the desired quality (or the best 
possible quality) of the spectrum of the new data series is achieved.

\subsection{The Extended Binomial Multifractal Model}

The multifractal cascade model \cite{feder88,barabasi,kantelhardt02} is a standard 
model for multifractal data, which is often applied, e.~g., in hydrology \cite{iturbe}.  
In the model, a record $x_i$ of length $N=2^{n_{\rm max}}$ is constructed recursively 
as follows. In generation $n=0$, the record elements are constant, i.~e.  $x_i = 1$ 
for all $i=1, \ldots, N$.  In the first step of the cascade (generation $n=1$), the 
first half of the series is multiplied by a factor $a$ and the second half of 
the series is multiplied by a factor $b$.  This yields $x_i=a$ for $i=1, \ldots, N/2$ 
and $x_i=b$ for $i=N/2+1, \ldots, N$. The parameters $a$ and $b$ are between zero 
and one, $0 < a < b < 1$. One need not restrict the model to $b = 1-a$ as is often 
done in the literature \cite{feder88}.  In the second step (generation $n=2$), we 
apply the process of step 1 to the two subseries, yielding $x_i=a^2$ for $i=1, 
\ldots, N/4$, $x_i=ab$ for $i=N/4+1, \ldots, N/2$, $x_i=ba=ab$ for $i=N/2+1, \ldots, 
3N/4$, and $x_i=b^2$ for $i=3N/4+1, \ldots, N$.  In general, in  step $n+1$, each 
subseries of step $n$ is divided into two subseries of equal length, and the first 
half of the $x_i$ is multiplied by $a$ while the second half is multiplied by $b$.  
For example, in generation $n=3$ the values in the eight subseries are $a^3, \; a^2b, 
\; a^2b, \; ab^2, \; a^2b, \; ab^2, \; ab^2, \;b^3$.  After $n_{\rm max}$ steps, the 
final generation has been reached, where all subseries have length 1 and no more 
splitting is possible.  We note that the final record can be written as $x_i = 
a^{n_{\rm max}-n(i-1)} b^{n(i-1)}$, where $n(i)$ is the number of digits 1 in the 
binary representation of the index $i$, e.~g. $n(13) = 3$, since 13 corresponds to 
binary 1101.

For this multiplicative cascade model, the formula for $\tau(q)$ has been 
derived earlier \cite{feder88,barabasi,kantelhardt02}.  The result is $\tau(q) = 
[-\ln(a^q + b^q) +q \ln(a+b)]/ \ln 2$ or
\begin{equation} h(q) = {1 \over q} - {\ln(a^q + b^q) \over q\ln 2} + 
{\ln(a+b) \over \ln 2}. \label{Hbin1} \end{equation}
It is easy to see that $h(1)=1$ for all values of $a$ and $b$.  Thus, in this form the 
model is limited to cases where $h(1)$, which is the exponent Hurst defined originally 
in the $R/S$ method, is equal to one.  

In order to generalise this multifractal cascade process such that any value of 
$h(1)$ is possible, one can subtract the offset $\Delta h = \ln(a+b) / \ln(2)$ from 
$h(q)$ \cite{jhydrol}.  The constant offset $\Delta h$ corresponds to additional 
long-term correlations incorporated in the multiplicative cascade model.  For 
generating records without this offset, we rescale the power spectrum.  First, we 
fast-Fourier transform (FFT) the simple multiplicative cascade data into the frequency 
domain.  Then, we multiply all Fourier coefficients by $f^{-\Delta h}$, where $f$ is 
the frequency.  This way, the slope $\beta$ of the power spectra $S(f) \sim 
f^{-\beta}$ is decreased from $\beta = 2 h(2) - 1 = [2 \ln(a+b) - \ln(a^2 + b^2)] 
/\ln 2$ into $\beta' = 2 [h(2) - \Delta h] - 1 = -\ln(a^2 + b^2) /\ln 2$.  Finally, 
backward FFT is employed to transform the signal back into the time domain.

\subsection{The Bi-fractal Model}

In some cases a simple bi-fractal model is already sufficient for modelling apparently
multifractal data \cite{jgra}.  For bi-fractal records the Renyi exponents $\tau(q)$ are 
characterised by two distinct slopes $\alpha_1$ and $\alpha_2$,
\begin{equation}
\tau(q) = \left \{ \begin{array}{ll}
q \alpha_1 - 1  \quad q \le q_\times\\
q \alpha_2 + q_\times(\alpha_1-\alpha_2) - 1  \quad q>q_\times\\
\end{array} \right.
\end{equation}
or
\begin{equation}
\tau(q) = \left \{ \begin{array}{ll}
q \alpha_1 + q_\times(\alpha_2-\alpha_1) - 1  \quad q \le q_\times\\
q \alpha_2 - 1  \quad q>q_\times\\
\end{array} \right. .
\end{equation}
If this behaviour is translated into the $h(q)$ picture using Eq.~(\ref{tauH}),
we obtain that $h(q)$ exhibits a plateau from $q = -\infty$ up to a certain
$q_\times$ and decays hyperbolically for $q > q_\times$,
\begin{equation}
h(q) = \left \{ \begin{array}{ll} \alpha_1  \quad q \le q_\times \\
q_\times(\alpha_1-\alpha_2) \frac{1}{q} + \alpha_2  \quad q>q_\times
\end{array} \label{bifracfit1} \right. ,
\end{equation}
or vice versa,
\begin{equation}
h(q) = \left \{ \begin{array}{ll}
q_\times(\alpha_2-\alpha_1) \frac{1}{q} + \alpha_1 \quad q \le q_\times \\
\alpha_2  \quad q>q_\times \end{array} \label{bifracfit2} \right. .
\end{equation}
Both versions of this bi-fractal model require three parameters. The multifractal 
spectrum is degenerated to two single points, thus its width can be defined as 
$\Delta \alpha=\alpha_1-\alpha_2$.

\section{Future Directions}

The most straightforward future direction is to analyse more types of time series
from other complex systems than those listed in Chapter 2 to check for the presence 
of fractal scaling and in particular long-term correlations.  Such applications may
include (i) data that are not recorded as function of time but as function of another
parameter and (ii) higher dimensional data.  In particular, the inter-relationship 
between fractal time series and spatially fractal structures can be studied.  Studies 
of {\it fields} with fractal scaling in time and space have already been performed 
in Geophysics.  In some cases studying new types of data will require 
dealing with more difficult types of non-stationarities and transient behaviour, 
making further development of the methods necessary.  In many studies, detrending
methods have not been applied yet.  However, discovering fractal scaling in more and 
more systems cannot be an aim on its own.

Up to now, the reasons for observed fractal or multifractal scaling are not clear 
in most applications.  It is thus highly desirable to study causes for fractal and 
multifractal correlations in time series, which is a difficult task, of course.
One approach might be based on modelling and comparing the fractal aspects of real
and modelled time series by applying the methods described in this article.  The 
fractal or multifractal characterisation can thus be helpful in improving the models.
For many applications, practically usable models which display fractal or transient 
fractal scaling still have to be developed.  One example for a model explaining 
fractal scaling might be a precipitation, storage, and runoff model, in which the 
fractal scaling of runoff time series could be explained by fractional integration 
of rainfall in soil, groundwater reservoirs, or river networks characterised by a 
fractal structure.  Also studies regarding the inter-relation between fractal 
scaling and complex networks, representing the structure of a complex system, are
desirable.  This way one could gain an interpretation of the causes for fractal 
behaviour.

Another direction of future research is regarding the linear and especially non-linear
inter-relations between several time series.  There is great need for improved methods 
characterising cross-correlations and similar statistical inter-relations between 
several non-stationary time series.  Most methods available so far are reserved to 
stationary data, which is, however, hardly found in natural recordings.  See 
\cite{stanley08} for a very recent approach analysing fractal cross-correlations
in non-stationary data. An even more
ambitious aim is the (time-dependent) characterisation of a larger network of signals.
In such a network, the signals themselves would represent the nodes, while the (possibly 
directed) inter-relations between each pair represent the links (or bonds) between the
nodes.  The properties of both, nodes and links can vary with time or change abruptly,
when the represented complex system goes through a phase transition.

Finally, more work will have to be invested in studying the practical consequences 
of fractal scaling in time series.  Studies should particularly focus on predictions
of future values and behaviour of time series and whole complex systems.  
This is very relevant, not only in hydrology and climate research, where a clear 
distinguishing of trends and natural fluctuations is crucial, but also for predicting 
dangerous medical events on-line in patients based on the continuous recording of 
time series.  \\

{\bf Acknowledgement:}
We thank Ronny Bartsch, Amir Bashan, Mikhail Bogachev, Armin Bunde, Jan Eichner, 
Shlomo Havlin, Diego Rybski, Aicko Schumann, and Stephan Zschiegner for helpful 
discussions. This work has been supported by the Deutsche Forschungsgemeinschaft 
(grant KA 1676/3) and the European Union (STREP project DAPHNet, grant 018474-2).

\noindent
\includegraphics[width=16.0cm]{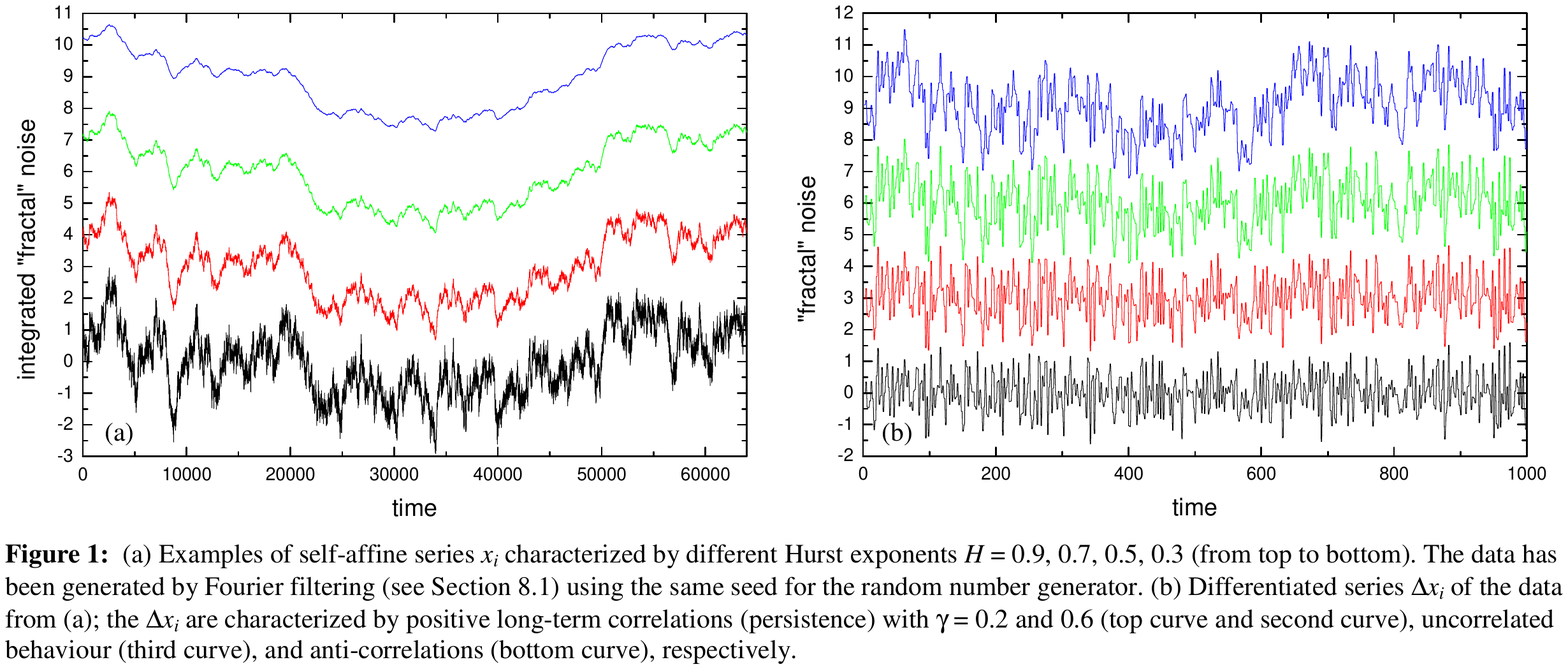}
\includegraphics[width=16.0cm]{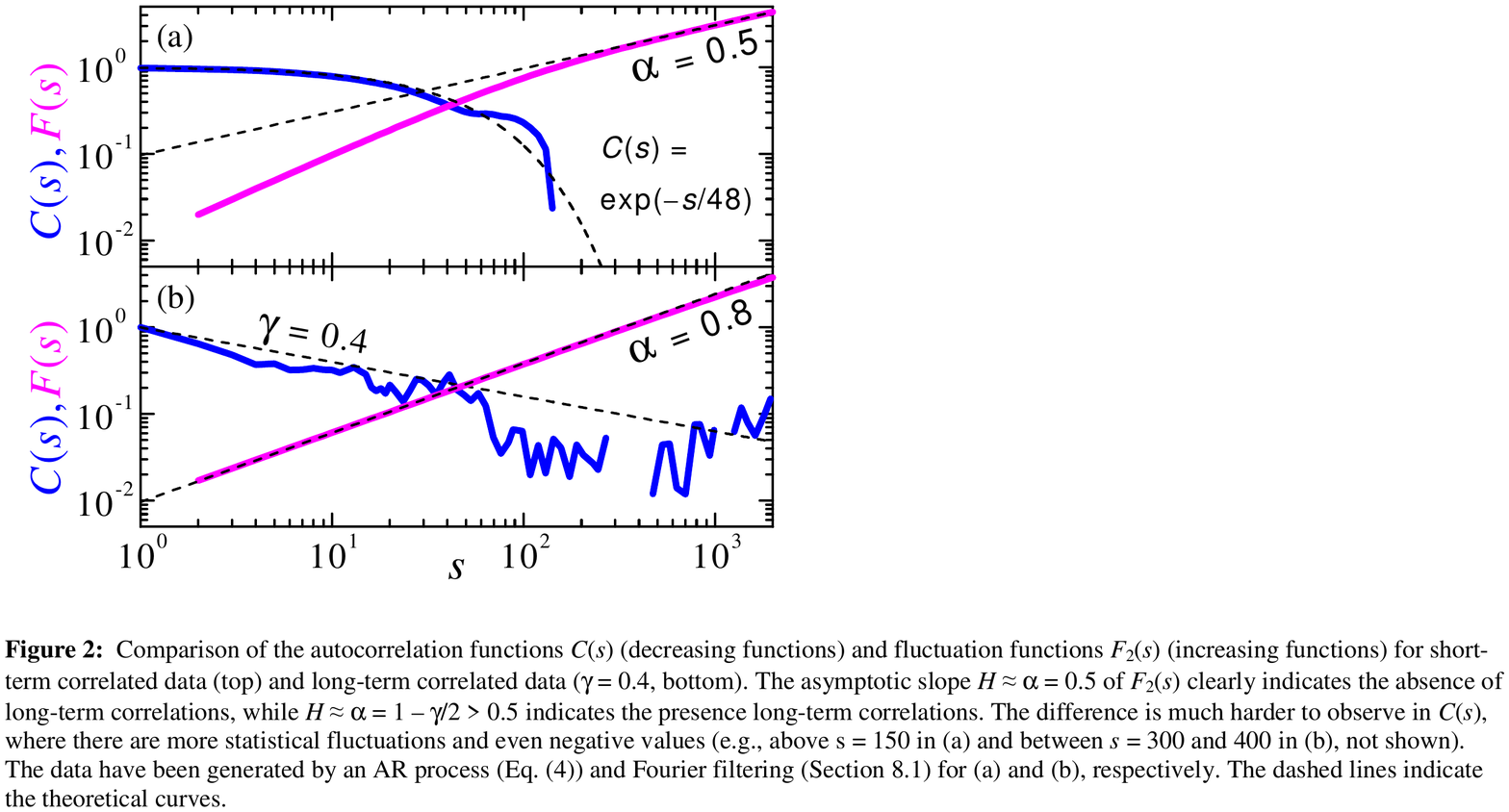}
\includegraphics[width=16.0cm]{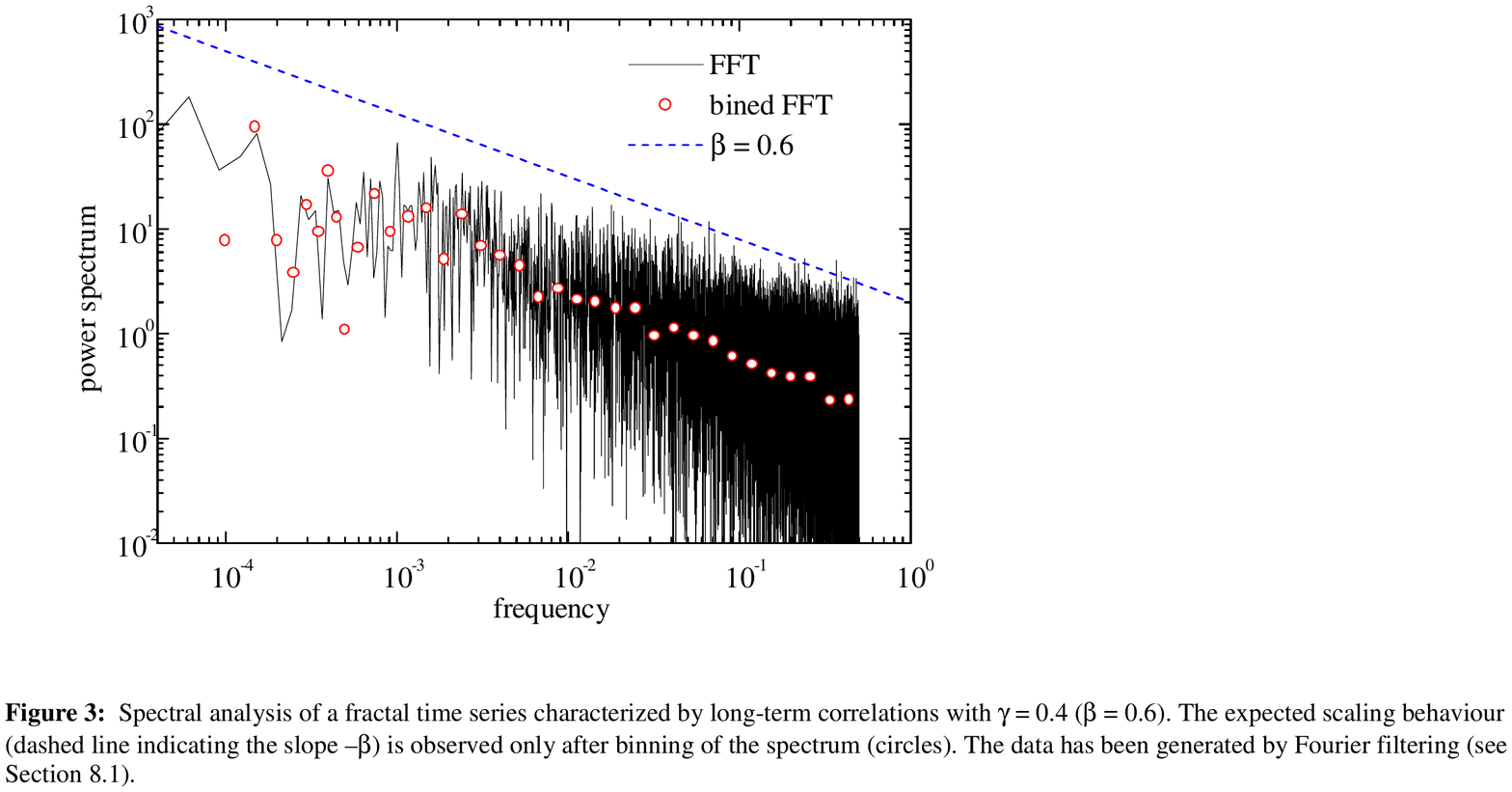}
\includegraphics[width=16.0cm]{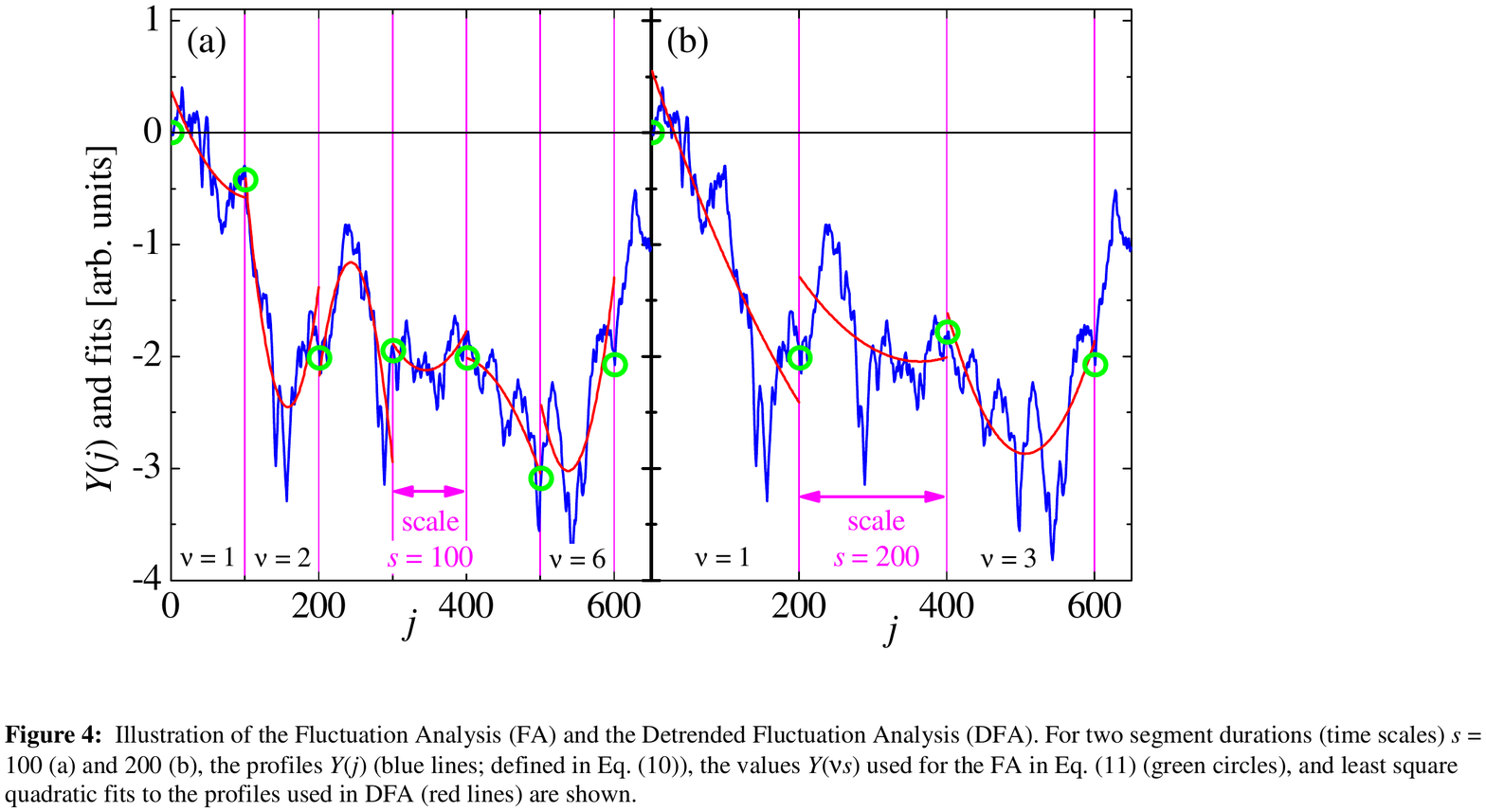}
\includegraphics[width=16.0cm]{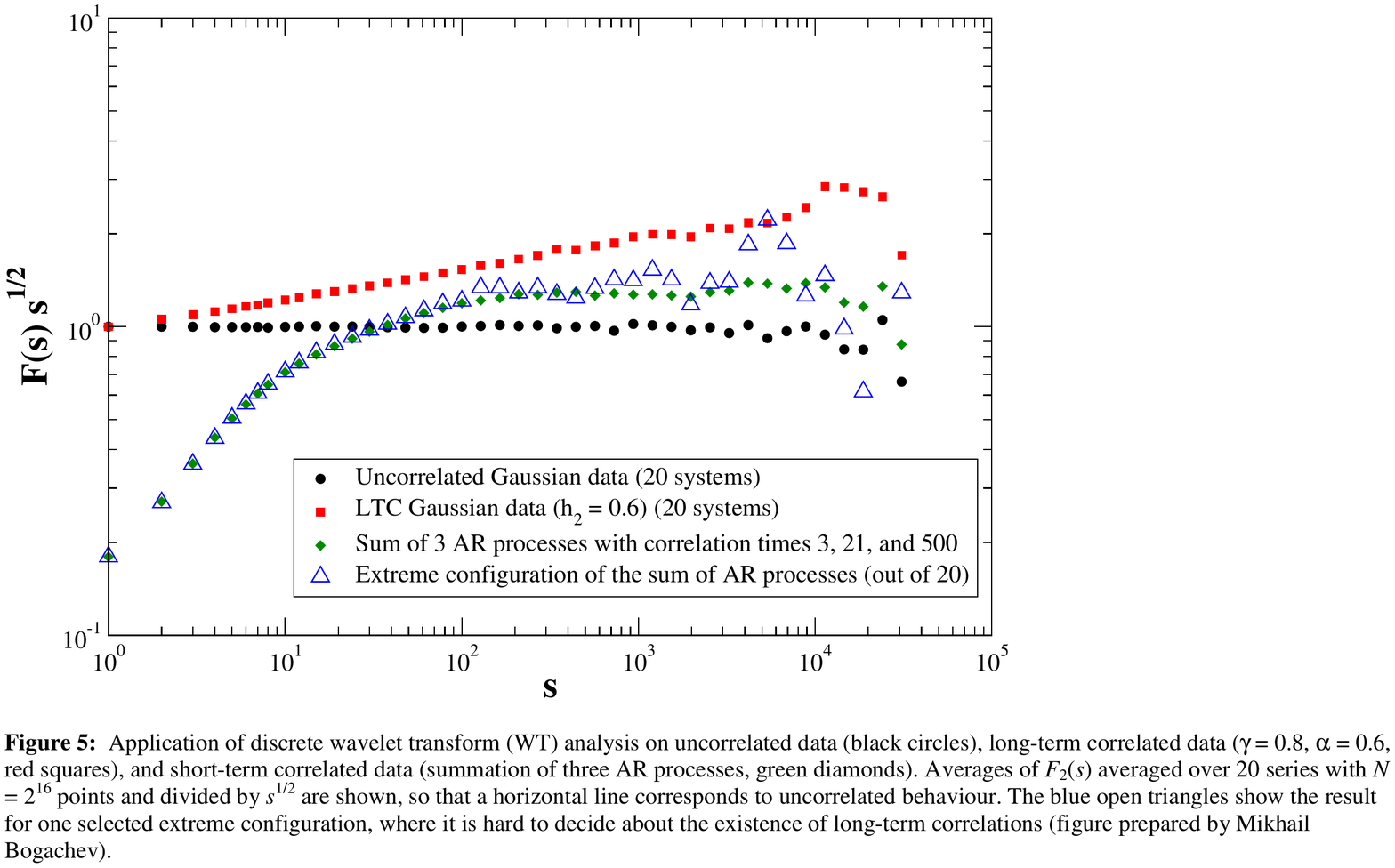}
\includegraphics[width=16.0cm]{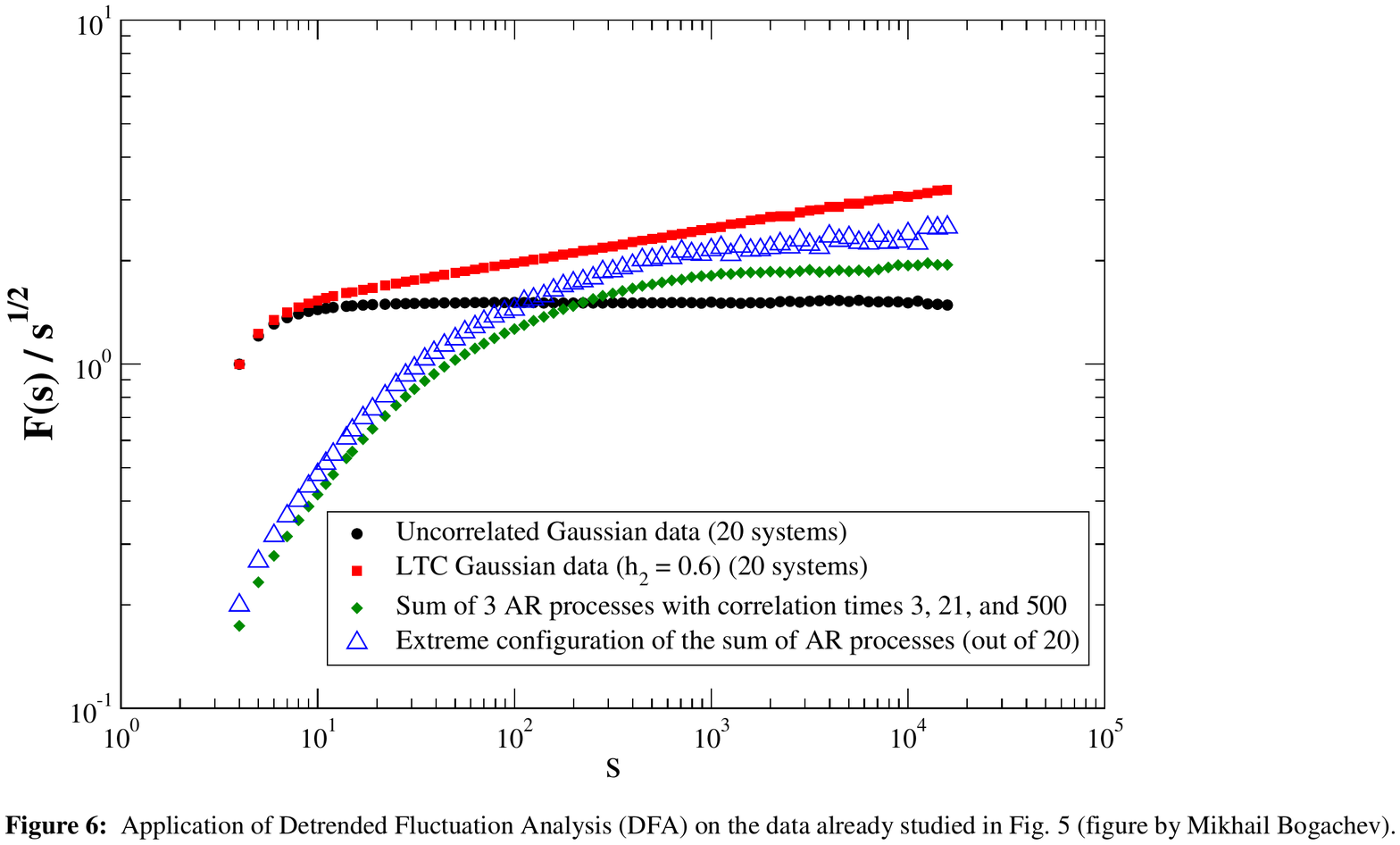}
\includegraphics[width=16.0cm]{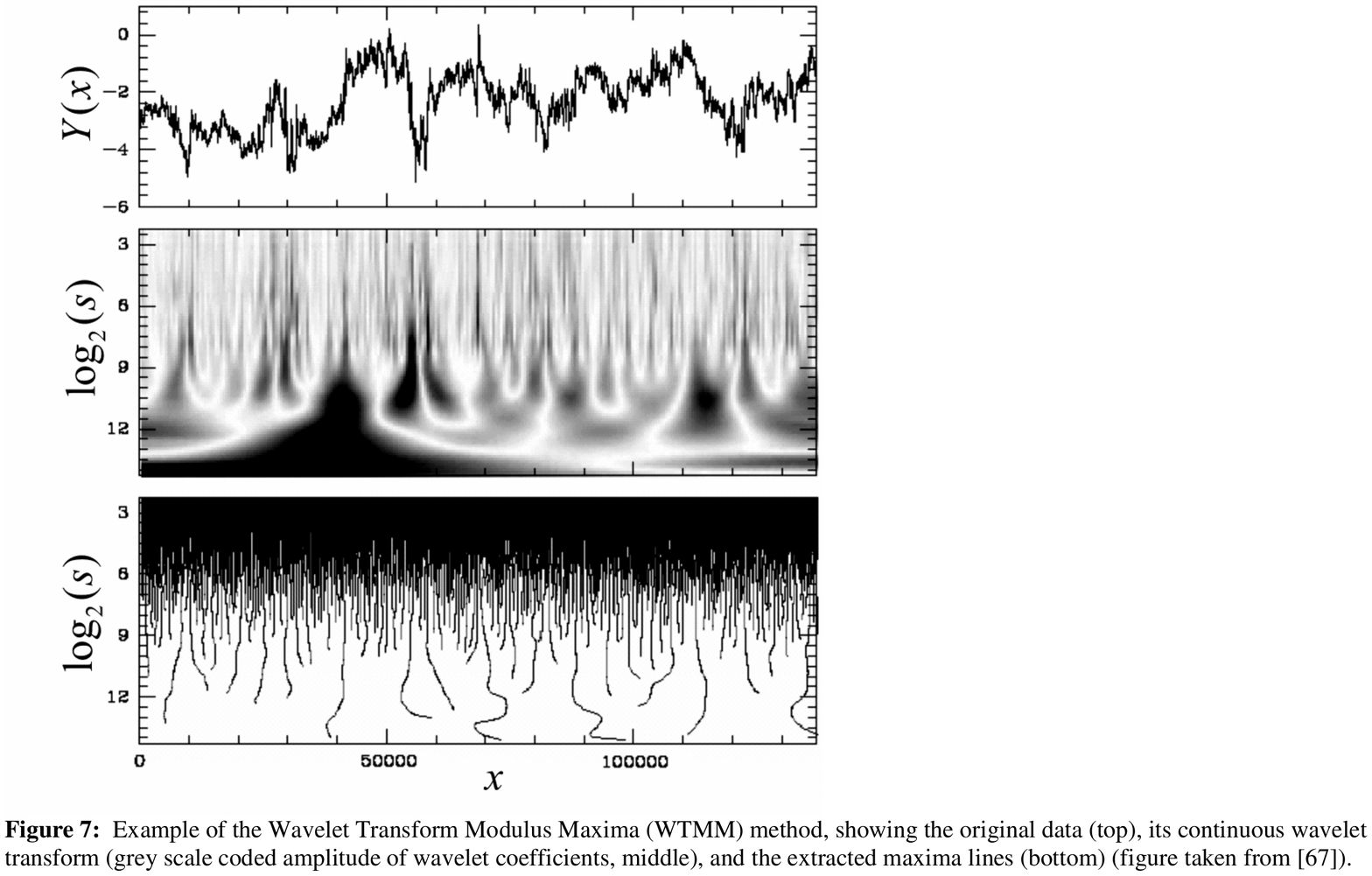}
\includegraphics[width=16.0cm]{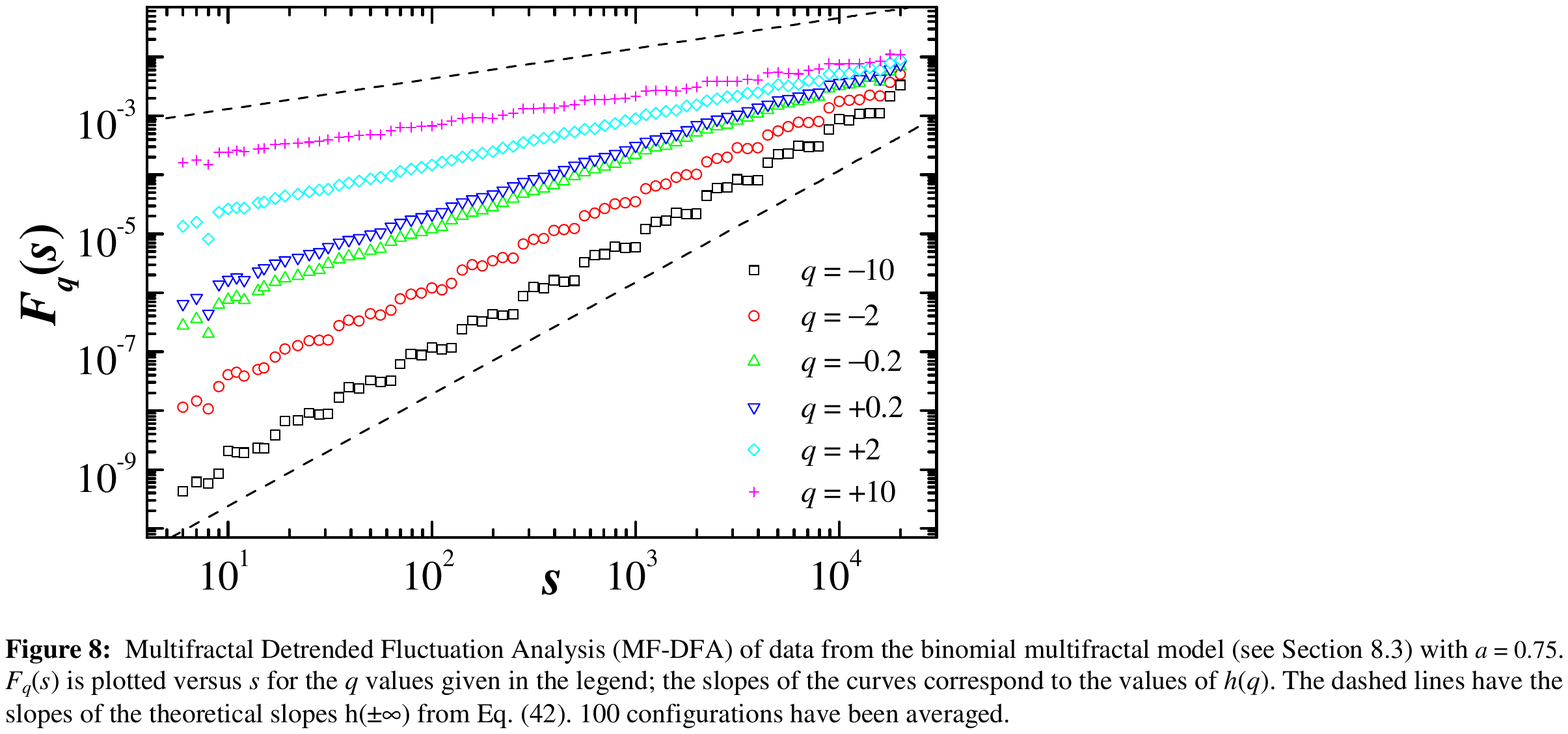}
\includegraphics[width=16.0cm]{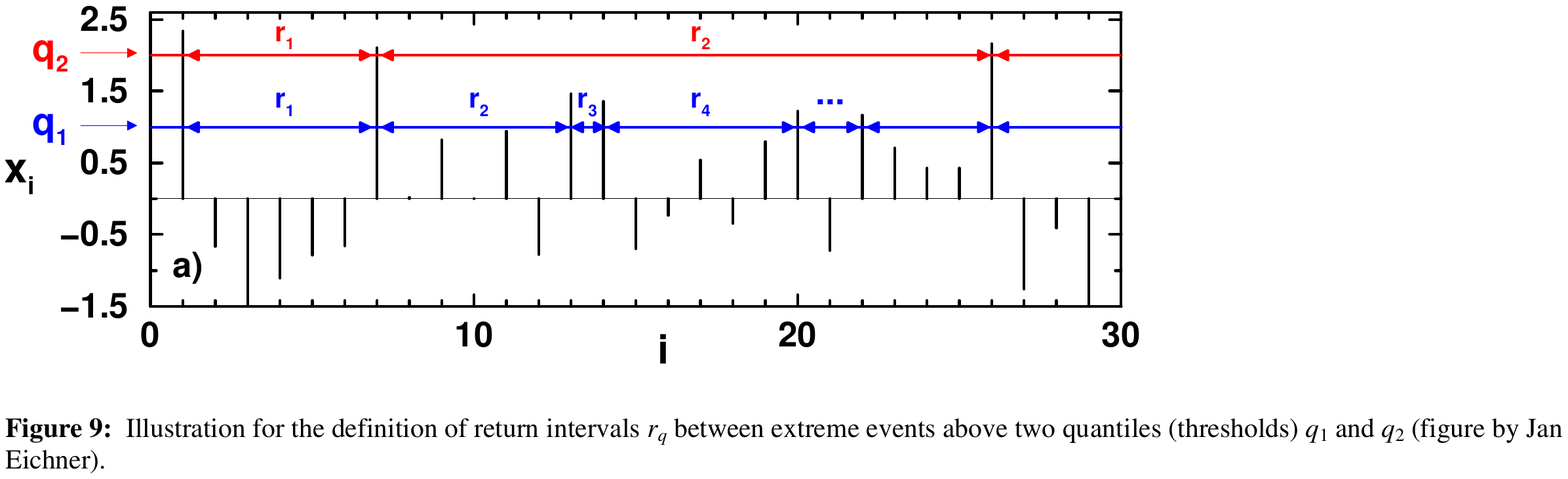}
\includegraphics[width=16.0cm]{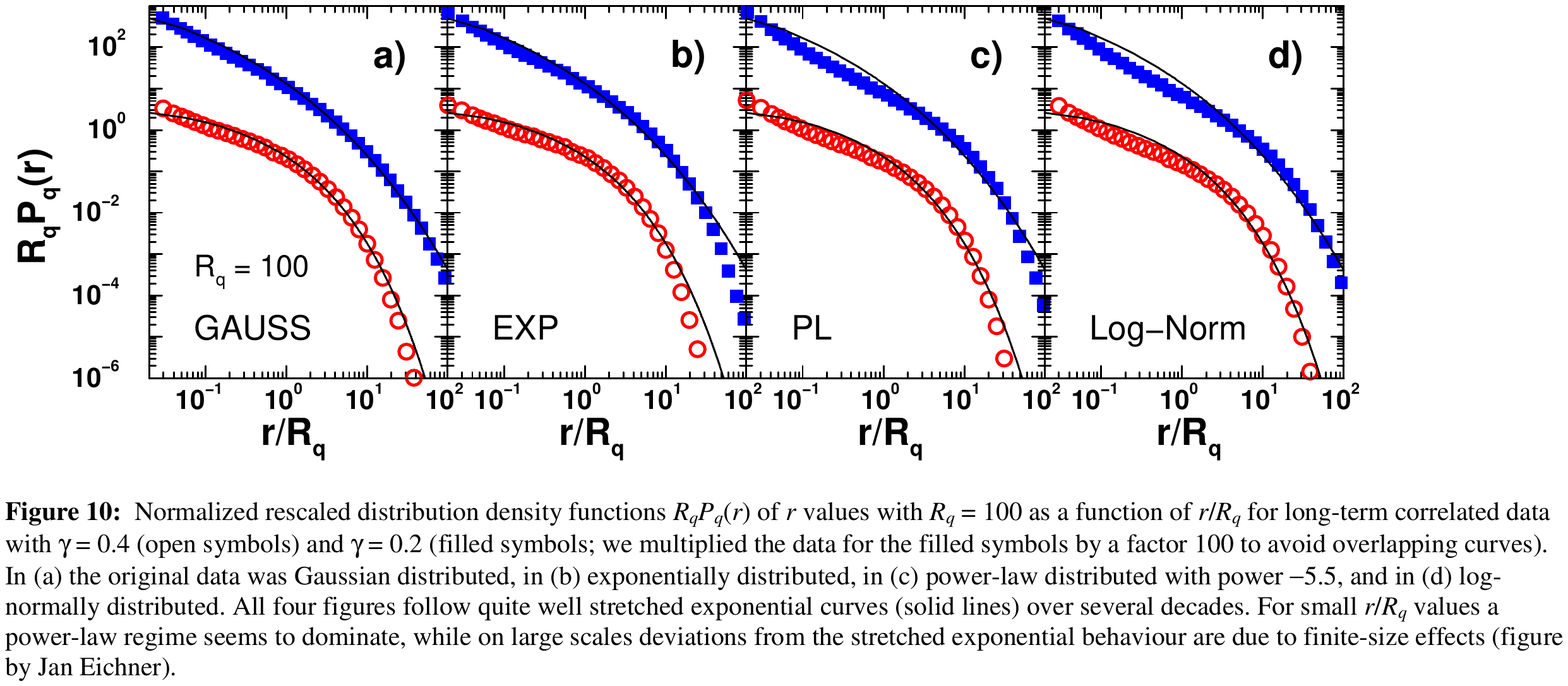}
\includegraphics[width=16.0cm]{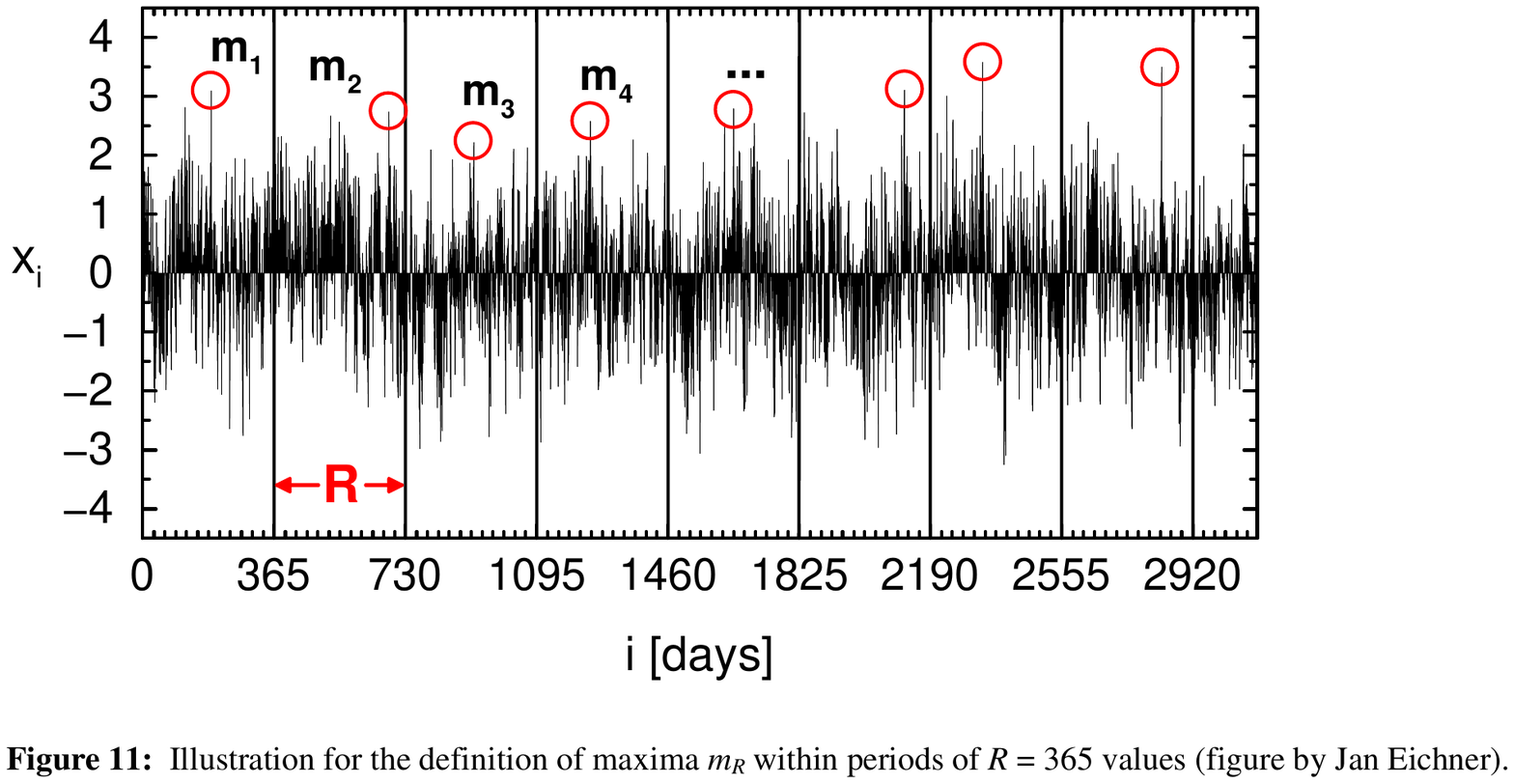}

\end{document}